# Metal–silicate partitioning of W and Mo and the role of carbon in controlling their abundances in the Bulk Silicate Earth

E.S. Jennings[a,b*], S.A. Jacobson[a,c,d], D.C. Rubie[a], Y. Nakajima[a,e], A.K. Vogel[a,f], L.A. Rose-Weston[a,g], D.J. Frost[a]

[a]Bayerisches Geoinstitut, University of Bayreuth, 95440 Bayreuth, Germany

[b]Department of Earth and Planetary Sciences, Birkbeck, University of London, Malet Street, London WC1E 7HX, United Kingdom

[c]Observatoire de la Côte d'Azur - Boulevard de l'Observatoire, CS 34229 - F 06304 NICE Cedex 4

[d]Department of Earth and Environmental Sciences, Michigan State University, East Lansing, MI, USA

[e]Present address: Department of Physics, Kumamoto University, 2-39-1 Kurokami, Kumamoto-shi, Kumamoto 860-8555, Japan

[f]Present address: Lava-Dome, Deutsches Vulkanmuseum Mendig, Brauerstr. 1, 56743 Mendig, Germany

[g]Present address: Barrick Gold Corporation - Hemlo Gold Mine, Marathon, Ontario, Canada.

*Corresponding author (e.jennings@bbk.ac.uk)

**Abstract**

The liquid metal–liquid silicate partitioning of molybdenum and tungsten during core formation must be well-constrained in order to understand the evolution of Earth and other planetary bodies, in particular because the Hf–W isotopic system is used to date early planetary evolution. The partition coefficients $D_{Mo}$ and $D_W$ have been suggested to depend on pressure, temperature, silicate and metal compositions, although previous studies have produced varying and inconsistent models. Additionally, the high cationic charges of W and Mo in silicate melts make their partition coefficients particularly sensitive to oxygen fugacity. We combine 48 new high pressure and temperature experimental results with a comprehensive database of previous experiments to re-examine the systematics of Mo and W partitioning, and produce revised partitioning models from the large combined dataset. W partitioning is particularly sensitive to silicate and metallic melt compositions and becomes more siderophile with increasing temperature. We show that W has a 6+ oxidation state in silicate melts over the full experimental $fO_2$ range of ΔIW -1.5 to -3.5. Mo has a 4+ oxidation state and its partitioning is less sensitive to silicate melt composition, but also depends on metallic melt composition. $D_{Mo}$ stays approximately constant with increasing depth in Earth. Both W and Mo become more siderophile with increasing C content of the metal: we therefore performed experiments with varying C concentrations and fit epsilon interaction parameters: $\varepsilon_C^{Mo}$ = -7.03 ± 0.30 and $\varepsilon_C^W$ = -7.38 ± 0.57.

W and Mo along with C are incorporated into a combined N-body accretion and core–mantle differentiation model, which already includes the major rock-forming elements as well as S, moderately and highly siderophile elements. In this model, oxidation and volatility gradients extend through the protoplanetary disk so that Earth accretes heterogeneously. These gradients, as well as the metal–silicate equilibration pressure, are fitted using a least squares optimisation so that the model Earth-like planet reproduces the composition of the Bulk Silicate Earth (BSE) across 17 simulated element concentrations (Mg, Fe, Si, Ni, Co, Nb, Ta, V, Cr, S, Pt, Pd, Ru, Ir, W, Mo, and C). The effects of the interaction of W and Mo with Si, S, O, and C in metal are included. Using this model with six separate terrestrial planet accretion simulations, we show that W and Mo require the early accreting Earth to be sulfur-depleted and carbon-enriched so that W and Mo are efficiently partitioned into Earth's core and do not accumulate in the mantle. If this is the case, the produced Earth-like planets possess mantle compositions matching the BSE for all simulated elements. However, there are two distinct groups of estimates of the bulk mantle's C abundance in the literature: low (~100 ppm), and high (~800 ppm), and all six models are consistent with the higher estimated carbon abundance. The low BSE C abundance would be achievable when the effects of the segregation of dispersed metal droplets produced in deep magma oceans by the disproportionation of $Fe^{2+}$ to $Fe^{3+}$ plus metallic Fe is considered.

1. Introduction

The accretion of Earth and segregation of its core are now widely regarded as being parts of a simultaneous, heterogeneous process. The composition of accreting impactors is likely to have changed through time with increased mixing in the protoplanetary disk, which would have resulted in changing oxygen fugacity and volatile contents throughout the formation of the early Earth (Wänke, 1981). Calculations and simulations suggest that large impacts release enough kinetic energy to cause widespread melting of the silicate Earth, resulting in magma ocean stages during its accretion (Melosh, 1990; Piarazzo et al., 1997; de Vries et al., 2016). Magma ocean formation would allow metal and silicate from an impacting body to equilibrate with at least a localised portion of the bulk silicate Earth (BSE). The distribution of siderophile elements, as estimated from the composition of the BSE, must trace this process. Different elements have partition coefficients which are affected by pressure, temperature, oxygen fugacity and melt composition in different ways (e.g. Mann et al., 2009; Siebert et al., 2011). Earth's accretionary history can therefore be understood by identifying the set of conditions that is compatible with all such elements. Such an approach is followed in the combined N-body accretion and core – mantle differentiation model of Rubie et al. (2015; 2016).

W and Mo are members of the same group (VIb) of the periodic table and will therefore display similar behaviour in geochemical systems. Being refractory, they should be present in the bulk Earth in chondritic proportions, and they are moderately siderophile, so measurable concentrations should remain in the silicate portion of the Earth following fractionation by the segregation of the core. In addition, W and Mo are both highly charged in silicate melts (O'Neill et al., 2008; Cottrell et al., 2009; Wade et al., 2012; 2013): because metal–silicate partitioning of siderophile elements usually involves a redox reaction, their high charges mean that their distribution will be particularly sensitive to the oxygen fugacity conditions during core segregation. However, their high charge also means that their activity in silicate melts is dependent on silicate melt composition (Thibault and Walter, 1995; Wade et al., 2012), and their activities in metallic melts are known to be highly dependent on the metallic melt's light element content (Jana and Walker, 1997a; Chabot et al., 2006; Cottrell et al., 2009; Righter et al., 2010). These dependencies on a variety of factors mean that W and Mo have the potential to be sensitive tracers of a range of aspects of core formation – especially oxygen fugacity and metal composition, but likewise mean that a large number of experiments are required to constrain the competing effects.

The Mo/W ratio in the BSE is 3.9 (Palme and O'Neill, 2014), which is low relative to the chondrite value of 10, and implies fractionation during core formation. Wade et al. (2012) found this low ratio difficult to reproduce with their partitioning data and a continuous accretion–differentiation model, which predicted that core formation should increase, not decrease, the mantle Mo/W ratio (for conditions that satisfy other siderophile elements): thus, they proposed that S in core-forming metal may resolve this. In a preliminary investigation, we added the parametrised partition coefficients of Wade et al. (2012) to the model of Rubie et al. (2015) (which also considers heterogeneous accretion but uses partial, rather than full, equilibration of the silicate Earth with the cores of impactors) and predicted a BSE that is overabundant in both Mo and W at conditions that satisfy the other elements considered, together with a Mo/W ratio that is low relative to that of the BSE.

Despite a large experimental effort by the community (section 1.1), there is some still uncertainty and disagreement regarding the partitioning behaviour of Mo and W between metal and silicate. The partitioning of these elements is not only important for core formation and accretion modelling: W partitioning needs to be well-constrained, given the importance of the short-lived $^{182}$Hf–$^{182}$W system in dating core formation and early planetary evolution of Earth, Moon and other solar system objects (Kleine et al., 2009). For example, a higher initial W content than that assumed from the present-day terrestrial mantle would require core-formation to end earlier in order to explain the ingrowth of radiogenic W.

In this study, we thoroughly constrain the partitioning behaviour of W and Mo in terms of valence (oxygen fugacity dependence), pressure and temperature dependencies of equilibration, and the composition of the metallic melts, by combining results from 48 new experiments with a larger database of all existing ones (subject to certain filters and quality control). The BSE concentrations of W and Mo are interpreted by adding these elements, together with C, to the combined accretion/core formation model of Rubie et al. (2015, 2016). With such a large experimental database, we are confident that the partitioning of these elements, including dependence on $P$, $T$, $fO_2$ and other compositional effects (using both new and recently-published interaction parameters), is now well-enough constrained to be used as a tool to investigate the processes of Earth's accretion.

*1.1. Previous studies and a big dataset approach*

Previous studies paint a somewhat contradictory picture of the $P$–$T$ partitioning systematics of W and Mo. Cottrell et al. (2009) found that increasing temperature slightly decreases the metal–silicate partition coefficient of W, whereas other studies (Righter et al., 2010; Siebert et al., 2011, Wade et al., 2012) have found the opposite. Likewise, the metal–silicate partition coefficient of Mo may either decrease (Righter et al., 2010; Siebert et al., 2011) or increase (Wade et al., 2013) with temperature at a fixed pressure. Whereas Righter et al. (1997) found a significant positive pressure dependence of the partitioning of W, Cottrell et al. (2009) determined that increasing pressure increases $D_W$ at low pressures (< ~ 4 GPa) but decreases it at higher pressures. Later, Wade et al. (2012) argued that the data are consistent with just a negative pressure dependence over the full pressure range. All studies, including ours, suffer from covariance of $P$ and $T$, i.e. higher temperatures are needed to melt the higher pressure experiments, so some disagreement is unsurprising. Whilst the impact of this uncertainty on accretion models is diminished by considering correlated $P$–$T$ profiles along a geotherm, these terms are still worth refining from a larger combined dataset in order to decrease uncertainty.

The compositions of both silicate and metallic melt phases are known to strongly affect the partitioning of W and Mo (Walter and Thibault, 1995; Righter and Drake, 1999; Righter et al., 1997, 2010; O'Neill and Eggins, 2002; Chabot et al., 2006; O'Neill et al., 2008; Siebert et al., 2011; Wade et al., 2012; Steenstra et al., 2016, 2017, 2020; sections 4.4 and 4.7 below). A potential cause of apparent discrepancies between previous studies is therefore the range in starting compositions and capsule materials used, which also affects the way in which experimental results should be applied to understanding planetary accretion. Many experiments have been performed in graphite capsules and have several wt.% carbon in the metal as a result (e.g. Jana and Walker, 1997a; Cottrell et al., 2009; Righter et al., 2010; Siebert et al., 2011): they yield partition coefficients for W and Mo that are around an order of magnitude higher than for C-free experiments, requiring activity corrections with large associated uncertainties for direct comparison with C-free experiments. Some studies used basaltic silicate compositions for ease of melting and quenching to a glass, whereas others use peridotitic compositions to emulate magma ocean compositions: different partition coefficients would be expected. Compositional controls can also be convoluted with the effect of temperature, because higher temperatures and/or longer experimental durations in MgO capsules will result in greater reaction between capsule and silicate, resulting in increased MgO concentrations and decreased concentrations of other components.

For this study, we have created new parameterisations of W and Mo partitioning based on a large number of experiments from many new and published studies, whereas previously published parameterisations are based on experiments from single laboratories or research groups. Using a compiled dataset approach means that we can interrogate pressure, temperature and melt composition effects simultaneously and more thoroughly than is possible with smaller datasets. In particular, including published data means that a wide range of silicate and metal compositions are considered at different conditions. This should reduce problems of covariance (though not eliminate them), allow fitting over a large parameter space (i.e. reduce the amount of extrapolation needed when modelling) and gives us the opportunity to filter the data to cover a narrower range of compositions that are relevant to magma oceans. There will also be inter-laboratory differences in pressure calibrations, temperature measurements, and analytical calibrations and techniques, that contribute to the offsets between studies: a large database approach smooths out these differences. From a statistical perspective, the more data that contributes to a parameterisation, the more robust that parameterisation will be.

## 2. Methods

We performed new experiments and re-analysed some previously published experimental data and combined these with a collated database of published experimental data to re-determine pressure, temperature and compositional dependencies of W and Mo partitioning. Within our new database, experiments were performed and analysed at different times by different protocols. Experiments which belong together as a set are indicated by the label under "group", and full experimental and analytical details as well as starting materials are documented in supplementary Table S1. All experiments were performed to examine the partitioning of W and/or Mo as well as other elements, except those labelled ESJ, which were performed to revise the dependencies of partitioning on the concentrations of C in Fe-rich liquid metal alloy.

### *2.1. Experimental methods*

Experiments were performed over a range of pressures and temperatures using piston-cylinder and multi-anvil apparatus at the Bayerisches Geoinstitut. Starting silicates had compositions that were either high-Mg basalt, simple primitive mantle, or olivine, and were prepared either using glassed and powdered high-purity oxide mixtures, or powdered synthetic or natural olivine. Metals were ground mixtures or ground sintered mixtures of high-purity metals. Materials were loaded as fine-grained mixtures or in the case of experiments labelled AV, as sandwiched layers (further details in supplementary Table S1). Starting materials were packed into MgO capsules in all experiments except A962b, for which a graphite capsule was used. Piston cylinder experiments were performed using a 1/2" talc-pyrex assembly with a graphite heater and type-D thermocouple using a hot-piston-in technique with a friction correction applied. Multi-anvil experiments used Cr-doped MgO octahedra positioned between truncated tungsten carbide anvils. Octahedra and truncation dimensions were appropriate to the pressure generated in each case. Assembly components and furnaces were heated at 1000 °C for several hours before use. Straight or stepped $LaCrO_3$ heaters were used, depending on the size of the assembly. Temperatures were either directly measured by a type-D thermocouple, or extrapolated along power-temperature curves. Uncertainties on pressure and temperature in the multi-anvil are around ± 0.5–1 GPa and ± 100 K, respectively. Details of starting materials and experimental conditions for each set of experiments can be found in supplementary Table S1 and Table 1.

*2.2. Analytical methods*

Samples were mounted, sectioned, polished, and subsequently analysed by electron probe microanalyser (EPMA) and/or laser-ablation inductively-coupled plasma mass spectrometry (LA-ICP-MS), both at the Bayerisches Geoinstitut. Major and minor elements in silicate and metal were measured by a JXA-8200 EPMA using wavelength-dispersive X-ray spectroscopy with a 15 kV beam. It should be noted that EPMA analyses of some multi-anvil experiments (labelled YN*) were previously published by Fischer et al. (2015), where further analytical details can be found. However, the LA-ICP-MS analyses of these samples are new.

A range of appropriate primary standards of metals, oxides, and silicates were used in the calibrations. LA-ICP-MS measurements were performed using a laser system which consists of an Elan DRC-e quadrupole mass spectrometer attached to a Geolas M 193 nm ArF Excimer Laser. The ablation cell was flushed with He to enhance sensitivity (Eggins et al., 1998; Günther and Heinrich, 1999). The concentration of S in the silicate melt, where present, was quantified using either Afghanite (Seo et al., 2011) or a S-bearing basaltic glass (SB19; Botcharnikov et al., 2011) standards. Ablation yields were standardised and drift was monitored with a NIST SRM 610 glass standard, and yields were normalised to EPMA Si (experiments labelled YN, AV and ESJ) or Ca (LRW) content. Spot sizes by EPMA and LA-ICP-MS in both silicate and metal were adjusted to account for the coarseness of the quench textures in order to obtain representative average compositions. In addition, a larger number of EPMA points were measured in the case of more problematic quench textures. Details of the analytical methods used for each set of experiments can be found in supplementary Table S1, Fischer et al. (2015), and Vogel et al. (2018). The resulting data are given in Table 2(a, b) and supplementary Table S2. Where the precision and accuracy of minor or trace elements was found to be similar for EPMA and LA-ICP-MS, the EPMA value is usually listed unless concentrations of that element are near or below the EPMA detection limit.

*2.2.1. Carbon analysis*

Experiments labelled 'ESJ' in Tables 1 and 2 included graphite powder in the starting mixture. Carbon partitions strongly into metal and will only be present in trace amounts in the silicate melt (Dasgupta and Walker, 2008; Chi et al., 2014; Duncan et al., 2017). The carbon content of the metal phase was measured in these experiments by EPMA. The measurement procedure was a modified version of that of Dasgupta and Walker (2008) with care taken to minimise and quantify carbon contamination, as described below.

A $Fe_3C$ primary standard was synthesised by inserting a 1 mm diameter, 10 mm long, 99.99 % purity Fe wire into a thick graphite sleeve and heating it to 1423 K at 15 kbar for one week. The reaction product was confirmed to have the $Fe_3C$ cementite structure by XRD and would be near-stoichiometric (Walker et al., 2013).

Organic C contamination prior to analysis was minimised by taking the following steps: 1) the epoxy sample mounts were left for one month and were then stored for 12 hours under a high vacuum to promote degassing; 2) samples and standards were subjected to N plasma cleaning immediately prior to analysis. For the metal analyses, neither the mounts nor the standards were carbon coated. Copper strips and Ag paint were used to create an electrical connection between the metal sample or standards and their respective metal holders. Analyses were performed at 15 kV and 25 nA using a 30 μm defocussed beam. The extent to which carbon drifts towards the beam was tested by measuring a continuous time series on a single spot on the pure Fe standard: it was found that over a period of ten minutes of continuous beam exposure, the increase in C counts was negligible, and the apparent decrease in C over the first ten seconds noted by Dasgupta and Walker (2008) was not observed (supplementary Fig. S1). Carbon was measured using a LDE2 crystal, and a pure Fe standard was measured between the analysis of each sample in order to monitor and quantify the carbon background. This background of 0.58 ± 0.09 wt. % C was constant throughout the analytical session, and was subtracted from the final measurements of the metals. Carbon was measured for 10 s with background measurements of 5 s, to further mitigate organic carbon contamination (Dasgupta and Walker, 2008). The C peak was very broad, and required a wide background interval. Final C contents were broadly similar to those expected from the weight ratio used in starting material preparation.

## 3. Definitions and data treatment

Partitioning of an element M with a valence *n* between liquid metal and silicate melt can be described by the exchange reaction:

[eq. 1]   $MO_{n/2} + \frac{n}{2} Fe = M + \frac{n}{2} FeO$

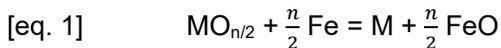

This reaction includes oxygen, and thus by expressing partitioning as an exchange reaction rather than an abundance ratio, the effect of $fO_2$ is implicitly accounted for. This is convenient, because $fO_2$ has a major effect on partitioning, and its determination is not without uncertainty. The equilibrium constant *K* can be split into an observed concentration ratio ($K_D$) and an activity correction:

[eq. 2] $$K = \frac{(a_M)(a_{FeO})^{\frac{n}{2}}}{(a_{MO_{n/2}})(a_{Fe})^{\frac{n}{2}}}$$

[eq. 3] $$K_D = \frac{(X_M)(X_{FeO})^{\frac{n}{2}}}{(X_{MO_{n/2}})(X_{Fe})^{\frac{n}{2}}}$$

[eq. 4] $$\log(K) = \log(K_D) + \log(\gamma_M) - \frac{n}{2}\log(\gamma_{Fe})$$

where $X$ always denotes mole fraction. In this study, we generally present findings in terms of $K$ (which can be considered as the activity-corrected observed $K_D$). We apply activity corrections for the metal compositions by multiplying $X$ by the activity coefficient $\gamma$, but neglect the silicate activity coefficients by assuming that their ratios remain constant. This approximation is useful for data presentation purposes in the absence of a well-constrained activity model for the components FeO, $WO_3$ and $MoO_2$ in silicate melts, and it is made in other studies (e.g. Wade and Wood, 2005). However, melt compositional effects identified by O'Neill and Eggins (2002) and Wade et al. (2012) indicate that this approximation may be over-simplistic for $WO_3$ and $MoO_2$: this point is further examined in section 4.4 below. $K$ can also be written in terms of the molar partition coefficient $D$ (eq. 5) and $fO_2$ (relative to the iron-wüstite buffer: ΔIW):

[eq. 5] $$D = \frac{X_M}{X_{MO_{n/2}}}$$

[eq. 6] $$fO_2(\Delta IW) = 2\log\left(\frac{a_{FeO}}{a_{Fe}}\right)$$

We follow the approach used by Wade and Wood (2005) of using $\varepsilon$ interaction parameters (Wagner, 1952) with the equations of Ma (2001) to quantify the activity terms for components in the metal ($\gamma_{Fe}$, $\gamma_M$). This procedure accounts for first-order interactions between the various solutes and iron in a way that is consistent with the Gibbs-Duhem equation. We calculate $\gamma_{Fe}$ using eq. 23 of Ma (2001) and $\gamma_M$ with their eq. 24. For these calculations, the composition of the metal is simplified by considering only the mole fractions ($X$) of Fe, Ni, S, O, Si, W and Mo, ± C. These are known major or minor components of both core-forming liquids and experimental metallic liquids, with mostly known and sometimes large mutual effects on the activities of the other components. $\varepsilon_i^j$ of each solute pair ($i$, $j$) is required, as well as the activity coefficient at infinite dilution of the solute of interest ($\gamma_M^0$; M = Mo or W). Experimental values of $\varepsilon_i^j$ and $\ln(\gamma_M^0)$ are published at a stated temperature ($T'$), and can be extrapolated to the temperature of each partitioning experiment ($T$) by multiplying by ($T'/T$) (J.S.P.S, 1988). J.S.P.S (1988) provides an extensive compilation of experimental values of $\varepsilon_i^j$

(given as $e_i^j$ and converted to $\varepsilon_i^j$ using their eq. 20) and $\gamma^0$. Values of $\varepsilon_i^j$ and $\gamma_M^0$ for several elements have recently been revised in high pressure studies with a focus on metal–silicate partitioning in the context of core formation (e.g. Tuff et al., 2011; Wood et al., 2014; Fischer et al., 2015). Here we use only values that were determined independently, i.e. that were not potentially affected by covariance with other fitting parameters. The values and references used in our activity corrections are listed in Table 3. The parameters $\varepsilon_C^W$ and $\varepsilon_C^{Mo}$ have been re-evaluated in the present study (see section 4.7).

## 4. Results

Major and trace element analyses of metal and silicate portions of the experimental samples are presented in Table 2(a, b), and calculated values of $fO_2$ (relative to the iron-wüstite buffer, ΔIW), $K_D$ and other parameters are listed in Table 4. The complete data, including trace elements omitted in Table 4, are provided in Supplementary Table S2. When EPMA and LA-ICP-MS are both used to analyse a given element, EPMA results are listed if uncertainties are smaller than the LA-ICP-MS results.

### 4.1. Sample textures and reaction with the capsule

The silicate portions of all samples (except A962b) contain ubiquitous quench crystals, requiring analysis by either LA-ICP-MS and/or a large number of EPMA analyses performed with a broad defocussed beam to determine a representative average composition. Experiments performed with basaltic starting compositions (labelled ESJ) contain spinifex olivine in a glass matrix, whereas those performed on ultramafic compositions and/or at very high temperatures often contained additional ferropericlase crystals and more complex, intricate quench textures (YN, AV, LRW) (Fig. 1). The elevated MgO content of the silicate relative to that of the starting materials indicates that reaction occurred with the capsule walls (Table 2a, b). The large euhedral ferropericlase crystals in high-$T$, high-MgO experiments are clearly texturally distinct from quench crystals (Fig. 1c) and grew during the experiment rather than precipitating during quench, saturating at the run temperature as melt MgO concentrations increased. Such crystals were therefore avoided when selecting analytical points.

The metal phase always coalesced into one or more large spheres. These spheres sometimes displayed an intricate texture (heterogeneity on the order of 5–10 μm; Fig. 1), caused by exsolution into two immiscible metals during the quench. Tiny crystals of $SiO_2$ were also occasionally observed. The textures suggest complete miscibility at experimental temperatures, so that an integrated composition is required: a defocussed electron beam and/or a large number of analytical points were used to obtain the metal or silicate bulk compositions. The experimental textures are similar to those reported in previous studies (e.g. Corgne et al., 2008, Siebert et al., 2011; Wade et al., 2012).

### 4.2. Attainment of equilibrium

Previously-published time-series experiments on a variety of elements show that equilibration times in metal–silicate partitioning experiments are generally very short, for example, 15 minutes or less at 1823 K, through to seconds to tens of seconds at 2300 K (Righter et al., 2010; Tuff et al., 2011; Thibault and Walter, 1995; Corgne et al., 2008). Specifically, Righter et al. (2010) demonstrated equilibration of W and Mo between metal and silicate within 30 minutes at 1873 K in the piston cylinder. We performed five time series experiments in a piston cylinder apparatus, and note that there is no resolvable or consistent change in $K_D^W$ at 2223 K when experimental times are increased from 15 to 100 minutes. Experimental durations (Table 1) were thus chosen to be significantly longer than the minimum required for equilibration (as guided by this and previous studies), yet short enough to minimise reaction with the capsule.

Most experiments appear well-equilibrated, based on the lack of change in compositions with distance from phase boundaries. An exception to this is group YN. In this group of experiments, ferropericlase precipitation was significant, especially on the metal margins, occasionally blocking diffusional pathways between metal and silicate. Sometimes, the silicate seems to have a higher W concentration when the pathway to the metal appears long or convoluted in the 2-dimensional section. This is difficult to explain: given that W was added as a metal powder, incomplete equilibration should lead to decreasing W concentrations in the silicate away from the interface. Nevertheless, the resultant concentrations are more variable than those of all other trace elements, which is suggestive of some W disequilibrium over long distances. This does not appear to affect any other elements. To account for possible disequilibrium, only analytical spots close to the metal, and those with unrestricted pathways to the metal, were used in determining silicate compositions from YN experiments.

### 4.3. Combination with published data

We fitted a predictive model for $K^{Mo}$ and $K^W$ to the experimental data. To do this, results from our new experiments were combined with an existing dataset of published silicate and metal compositions in partitioning experiments, to maximise the P–T–X range conditions considered. We compiled a large database of all known partitioning experiments with Mo and/or W partitioning measurements, and applied filters to the data. For consistency, we recalculated mole fractions, $K_D$ and activity corrections, so experiments for which the complete metal and silicate compositions are not published could not be used. For the C-free fits, we do not include experiments that were performed in graphite capsules. Studies with experiments that were considered are: W (n = 118): Hillgren et al. (1996); Jana and Walker (1997a); Kilburn and Wood (1997); Righter and Drake (1999); Holzheid and Palme (2007); Cottrell et al. (2009); Righter et al (2010); Seclaman (2016); Shofner (2011); Siebert et al. (2011); Tuff et al. (2011); Wade et al. (2012); Wood et al. (2014). Mo (n = 112): Hillgren et al. (1996); Jana and Walker (1997a); Righter et al. (1997); Righter and Drake (1999); Holzheid and Palme (2007); Righter et al. (2010); Siebert et al. (2011); Shofner (2011); Tuff et al. (2011); Wade et al. (2012); Burkemper (2013); Wood et al. (2014). We have not used data from

Steenstra et al. (2017) because their unusual silicate starting compositions mean that the partition coefficients are likely to be different to those relevant to terrestrial magma ocean fractionation (see below).

If the same experimental result appears in multiple publications, only the first occurrence is cited, unless the sample was subsequently re-analysed. Some experimental data include concentrations of Mo and W that are too low to measure effectively in the silicate melt, which would add uncertainty to the model fits, so we only fitted data where the silicate contained > 1 ppm of trace elements and > 0.5 wt. % FeO in the silicate melt. In addition, we restrict experiments with metal compositions to those in which Mo or W are considered trace solutes in molten iron. The concentration limits for which the epsilon model is appropriate are unclear, but a metal saturated in S or C should no longer be described in terms of interactions between trace solutes and iron. We cautiously limited the metal composition to C-free, $X_S < 0.1$ (to minimise uncertainty from interactions with S), $X_{Mo}$ or $X_W < 0.1$ (self-interactions are not well constrained), and $X_{Fe} > 0.75$. With the data filtered, 108 W-bearing experiments and 82 Mo-bearing experiments remained.

### 4.4. Effects of silicate melt composition on partitioning

Walter and Thibault (1995) first suggested, from the perspective of melt structure, that silicate melt composition could affect Mo and W partitioning. $W^{6+}$ is a large and highly-charged cation, so the polymerisation of the melt or the availability of large, highly-coordinated sites would influence its activity. This manifests as an increase in $D$ and $K_D$ with increasing melt polymerisation, as the incorporation of $WO_3$ becomes less favourable. Experiments indicate that increased melt polymerisation also reduces $MoO_2$ solubility (Jana and Walker, 1997b; Burkemper et al., 2012) in silicate melts, although the effect should be smaller than for tungsten. The effects of silicate melt composition on $D_W$ and $D_{Mo}$ were recently examined by Steenstra et al. (2017), who experimented with a suite of unusual compositions and identified compositional dependencies of $D_W$ and $D_{Mo}$; in particular, they identified a strong negative correlation with melt CaO concentration, consistent with the finding of O'Neill and Eggins (2002) that $\gamma_{MoO_2}$ has a strong dependence on silicate CaO,

$K$ is calculated by assuming a constant activity ratio between FeO and $WO_3$ or $MoO_2$, so it will show an apparent change if the oxide activity coefficients change. The effect of silicate melt composition can therefore be seen by plotting $K$ as a function of various melt compositional parameters (Fig. 2). Because $D_W$ is sensitive to temperature but not to pressure (section 4.5), experiments are grouped by temperature in Fig. 2. It should be noted that the melt composition is to some extent correlated with temperature (higher temperatures are needed to reach the peridotite liquidus than the basalt liquidus, and reaction with the MgO capsule occurs more rapidly at high temperatures), so the temperature and compositional effects on partitioning are convoluted. There is a small resolvable dependence of $K$ on composition within a given temperature range and a stronger trend is observed when considering all data, indicating that much of the apparent compositional dependence is really a temperature effect. However, there do appear to be real compositional dependencies within the narrower temperature brackets, highlighting some non-ideal interactions. Two causes are proposed in the literature, and the data are consistent with both: 1) $K^W$ positively correlates with $Al_2O_3$ and $SiO_2$, suggesting a melt polymerisation control (Walter and Thibault, 1995; Jana and Walker, 1997b), and 2) $K^W$ correlates negatively with CaO, suggesting that the formation of Ca-W complexes in the silicate melt reduces the activity of W (O'Neill and Eggins, 2002; O'Neill et al., 2008; Steenstra et al., 2017).

The effect of silicate composition on $K^{Mo}$ in this data compilation is not resolvable (except for some anomalously high $K^{Mo}$ values at > 2 wt. % $Na_2O$; Supplementary Fig. S2). This is in contrast to the findings of O'Neill and Eggins (2002), who identified a strong silicate melt control (especially CaO concentration) on $\gamma_{MoO2}$ when examining a broader compositional range in a study specifically designed to constrain $\gamma_{MoO2}$, i.e. other complicating factors were minimised. There may, in fact, be some silicate compositional effect on $K^{Mo}$, but because of the competing effects of different aspects of melt composition, different experimental setups and relatively narrow compositional range included here, this effect is not resolvable in our dataset within uncertainty. For this study, we therefore assume that the $MoO_2$ and FeO activity coefficient ratios in silicate melts are approximately constant over the range of compositions appropriate to core formation conditions.

Previous studies have included an activity correction for the silicate melt composition by: (1) fitting an NBO/T term as a proxy for silicate melt effects (Walter and Thibault, 1995; Righter et al., 1997; Siebert et al., 2011); (2) fitting a selection of single component terms but with little thermodynamic justification (e.g. Righter and Drake, 1999; Righter et al., 2010; Steenstra et al., 2016; Steenstra et al., 2017); and (3) fitting a regular solution model (e.g. Wade et al., 2012). NBO/T cannot encompass the full melt compositional effects for $MoO_2$ or $WO_3$ (O'Neill and Eggins, 2002; O'Neill et al., 2008) and it may not correctly describe melt structure at high pressure. The other approaches involve the simultaneous fitting of numerous highly-correlated terms which will be convoluted with both one-another and with the effect of temperature. We instead chose to restrict the silicate melt compositions to those relevant to early Earth magma oceans, i.e. broadly pyrolite-like to picritic, and check that there are no resolvable correlations between residuals in the fitted models and major element oxides in the silicate. Fig. 2 presents $K^W$ with the ranges in other oxides restricted, in order to better isolate the effect of each component. The final major element limits for the silicate compositions were chosen to minimise the compositional effect and are given in Table 5. Restricting the silicate MgO concentration will reduce the problem of correlation with temperature to some extent. After filtering data for silicate composition, 57 W-bearing experiments W and 68 Mo-bearing experiments remain (details in Supplementary Table S3).

### 4.5. P and T controls on partitioning

After filtering the data according to the silicate compositional limits given in Table 5, 57 experiments (29 from this study and 28 previously published) remained for fitting $K^W$ and 68 (13 this study; 55 published) for $K^{Mo}$. Pressure and temperature dependencies for $K^i$ were parametrised in the form:

[eq. 7]    $\log K^i = a + b/T + cP/T$

as proposed by Righter et al. (1997) and used in numerous previous studies. Because $K$ relates to $\Delta G°$, the terms a, b and c relate to $\Delta S°/R$, $\Delta H°/R$ and $\Delta V°/R$ respectively, although the ratio of silicate melt activity terms also accounts for some portion of the fitted parameters in eq. 7. Compositional effects in the metal phase are accounted for by the activity terms in eq. 4. Fitted terms (except the intercept) with a p-value of < 0.05 are considered significant and the goodness of fit of the model to the data is expressed by the root mean square (RMS) error. Because the reporting of uncertainties is inconsistent between different studies, fitting is not weighted according to uncertainty (although reported uncertainties are shown as error bars on figures). It should be noted that pressure and temperature are correlated parameters: as such, pressure and temperature dependencies cannot be interpreted independently when fitted together. Variance-covariance matrices are provided in Supplementary Table S4.

$\log K^W$: The pressure term (c) is statistically insignificant, so has a value of 0. Higher temperatures make W more siderophile over the experimental range, as is shown by a significant negative temperature term (b) (Fig. 3a). The temperature term is somewhat smaller than that of Wade et al. (2012; where b = -6728):

$\log K^W$: a = 0.61 (0.28); b = -4091 (670).    RMS = 0.311

$\log K^{Mo}$: Pressure and temperature terms (b and c) are small but statistically significant:

$\log K^{Mo}$: a = 1.47 (0.44); b = -1448 (851); c = -67.1 (20.9). RMS = 0.279

$K^{Mo}$ increases with temperature but decreases with pressure (Fig. 3b). From examining the experimental dataset, the net effect appears to be that there is little change in $K^{Mo}$ with depth in the Earth over the temperature and pressure range of experiments (1 atm to 25 GPa), where $\log K^{Mo}$ is approximately constant at 0.61 ± 0.30 (Fig. 3b). However, the pressure and temperature terms do have a more noticeable effect when $\log K^{Mo}$ is calculated for the more extreme conditions that may be relevant to magma oceans. The terms b and c are, respectively, larger and smaller than those determined by Wade et al. (2012) (where b = -143 and c = -167), and b has the opposite sign of that determined by Siebert et al. (2011) (where b = 3117).

The fitted a, b ± c terms are compared to published parameterisations of Wade et al. (2012), Cottrell et al. (2009), Righter et al. (2016), and Siebert et al. (2011) at 2 GPa and 20 GPa in Supplementary Fig. S3. The models are all superficially similar, though some have opposite signs for pressure and/or temperature effects. There is more disagreement in overall pressure and temperature-driven changes in $K_D^W$ than in $K_D^{Mo}$. Qualitatitive similarities are present with the pressure effects of the models of Righter et al. (2011) and Wade et al. (2012), who both find that $K^W$ has only a small (but statistically significant) negative pressure dependence, whereas $K^{Mo}$ has a larger one. Our models for $K^{Mo}$ and $K^W$ are most similar to those of Wade et al. (2012) for pyrolytic mantle, although are not identical. We note that the a, b and c terms of Wade et al. (2012) are extrapolated to pyrolytic compositions using a regular solution model, fitted to both their experiments and previous experiments available to the authors at the time: this smaller dataset was biased towards basaltic silicate compositions and featured many carbon-saturated experiments. It is not known whether the ratios of activity coefficients of W and Mo would change differently with pressure and temperature for these different compositions. Our approach represents the most cautious one in terms of limiting composition effects: our regression is by far the strictest of those published in terms of metal compositions, excluding high-C, high-S and high-Mo or W metal compositions, and considers a limited silicate compositional range.

There is no resolvable correlation between the residual errors and oxide concentrations in the silicate for $K^{Mo}$ or $K^W$. The model fits to the data are shown in Fig. 3. For both W and Mo, the full range of observed D values are well-predicted by the model (Fig. 4). We note that the highest pressure experiments in the calibration dataset are at 25 GPa, so these expressions (and previously published ones) may lose accuracy when extrapolated to higher pressures.

*4.6. Valence*

A good constraint on the valence of the oxide species is crucial for predicting the partitioning behaviour of Mo and W during core formation, by defining the dependence of $D_M$ on $fO_2$. By studying the solubility of W, Ertel et al. (1996) determined that $W^{4+}$ was the stable cation in silicate melts over their entire experimental $fO_2$ range (ΔIW -3.9 to -0.5). Later, Cottrell et al. (2009) used metal–silicate partitioning data to suggest that W was present as $W^{6+}$ at low pressures, also under reducing conditions, but that $W^{4+}$ stabilises at pressures greater than 11 GPa. Oxidation states from partitioning data were revisited and reviewed by Wade et al. (2012, 2013), who conclude that W exists in the 6+ state and Mo at 4+ up to at least 24 GPa, even at very low $fO_2$. Most recently, $W^{6+}$ was also suggested from partitioning data over a wide $fO_2$ range by Steenstra et al. (2020).

Oxidation states in silicate melts were measured directly using XANES for W and Mo by O'Neill et al (2008) and Righter et al. (2016), who identified $W^{6+}$ and $Mo^{4+}$ and respectively. Righter et al. (2016) found that $Mo^{4+}$ oxidised to $Mo^{6+}$ at around ΔIW > -1, i.e. at more oxidising conditions than those relevant to core formation conditions and present during most high pressure experiments.

The present study confirms those more recent conclusions: $W^{6+}$ and $Mo^{4+}$ are the stable species throughout the entire experimental range of pressures and oxygen fugacities. $D_M$ (molar ratio) is dependent on $fO_2$, and the two are related by rearranging eqs. 4, 6 and 7:

[eq. 8] $\quad \log(D_M) + \log(\gamma_M) - a - \frac{b}{T} - \frac{cP}{T} = \frac{-n}{4} fO_2 (\Delta IW)$

Oxygen fugacity $fO_2$ (ΔIW) is calculated from Fe and FeO activities in the respective metal and silicate phases using the activity coefficient $\gamma_{Fe}$, as determined by the epsilon model. No activity correction is applied to FeO in the silicate melt.

When the summed terms on the left side of eq. 8 are plotted against $fO_2$, the slope will be (-n/4) and the intercept should be zero. Plotting the data in this way removes compositional and P–T effects on partitioning (Fig. 5). To avoid circularity (the gradient of the trend is mostly independent of the data treatment, but the a, b and c terms will introduce a small bias towards the valence $n$ that was assumed when calculating K; eq. 2 and 7), we show the results obtained by separately fitting a, b and c for the valence states of +2, +4 and +6, and consider which produces the best quality fit (similar to Vogel et al., 2018). For $D_W$, using the a and b terms obtained with $n$ = 6, a line fitted through the data has a slope of -1.49 and yields a formal valence of 5.95 (root mean square RMS mismatch = 0.31). This result is robust regardless of $n$ assumed during fitting: if the data are plotted using the a and b terms obtained with $n$ = 4 yields almost the same gradient, corresponding to a formal valence of 5.50 (RMS = 0.35). The expected gradients corresponding to different valence states are shown on Fig. 5, and it is clear that, regardless of the valence assumed during fitting, the data still yield a steep gradient consistent with a high valence state of +6. Likewise for $D_{Mo}$, a formal valence ranging from 3.6 to 4.6 is obtained over the range of $n$ values (2 to 6) assumed during fitting: the best fit (lowest RMS) of $n$ = 4.12 is obtained when taking $n$ = 4 for fitting eq. 7. Again, the data support a 4+ valence state over the entire $fO_2$ range represented by the data. A lack of a systematic offset between predicted and actual values at extremes of $fO_2$ or extremes of D (figs. 4 and 5) are good evidence that the valence state does not change over the $fO_2$ range considered.

*4.7. C–W and C–Mo interactions*

It is well-known that C in metal can have a strong influence on the activities of other species, and that Mo and W are more siderophile when the metal is C-saturated (J.S.P.S. 1988; Jana and Walker, 1997a; Chabot et al., 2006; Cottrell et al., 2009; Righter et al., 2010). An increase in $D_M$ by around an order of magnitude in graphite-saturated experiments relative to C-free experiments can be seen in Fig. 6. The Earth's core probably contains a moderate fraction of C (e.g. McDonough 2003, Dasgupta and Walker 2008; Wood et al., 2013). However, because no metal–silicate partitioning experiments with variable carbon contents are published, the strength of interaction between C and Mo or W ($\varepsilon_C^M$) when C is undersaturated is not well quantified. A subset of experiments (labelled ESJ in Tables 1 and 2) were performed at 1923 K and 1.5 GPa to constrain the effect of C on partitioning. The four experiments in MgO capsules have nearly identical major element compositions and were equilibrated at identical P-T conditions, so any differences in $K_D$ are explained by the influence of metallic C on the Mo and W activities. These experiments show that both $\log(K_D^{Mo})$ and $\log(K_D^W)$ increase linearly with C content $X_C$ (Fig. 7), and the slope can be fitted with revised epsilon interaction parameters of $\varepsilon_C^{Mo}$ = -7.03 ± 0.30 and $\varepsilon_C^W$ = -7.38 ± 0.57 at a reference temperature of 1873 K. These values are similar to those listed in the Steelmaking Sourcebook (-6.03 and -6.45, respectively).

In fig. 7, the result of an experiment performed in a graphite capsule at the same conditions as those of the other ESJ experiments is also shown. $K_D^M$ is higher than predicted by the linear trend. This may be due to the effects of the lower MgO content of the silicate melt and higher content of other lithophile elements, as the starting composition has not been diluted by MgO derived from the capsule (Fig. 2). Alternatively, the interaction may no longer be linear at C-saturation and is not well-described with only first-order interaction parameters and the present activity model (indeed, the model of Ma, 2001, describes interactions between dilute components and is not intended for saturated systems). The latter explanation is consistent with solid-liquid iron partitioning data on many trace elements, where log($D$) values (as a function of C content) often begin to deviate from a linear trend at around 3–4 wt.% C (Chabot et al., 2006). The non-linearity indicates that correcting for carbon interaction in graphite-capsule experiments using the epsilon model (e.g. Cottrell et al., 2009) may result in the activity correction being underestimated. Our new $\varepsilon_C^{Mo}$ and $\varepsilon_C^W$ values can be used to predict the effect of up to around 3 wt. % C in the core-forming metal on the metal–silicate partitioning of siderophile elements.

*4.8. Interaction with carbon: molybdenum and tungsten compared with other siderophile elements*

Metallic Mo and W interact negatively and strongly with C which results in a significant increase in their metal–silicate partition coefficients with increasing C content (Fig. 7). A moderately negative C interaction parameter $\varepsilon_C^M$ is also listed for Cr (-4.9) in the J.S.P.S. (1988) compilation. By comparison, a weak positive interaction is listed for most other moderately siderophile elements: values for Co (1.7) and Ni (2.3) are small and positive. Increasing the C content of the metal should therefore slightly reduce the siderophility of Co and Ni. Chabot et al. (2005) observed no effect of C on the partitioning of these elements at 7 GPa between metal and silicate, although the solid-liquid metal partitioning data of Chabot et al. (2006) indicate that Ni and Co are indeed incorporated less favourably into C-bearing liquid iron. Volatile, moderately siderophile elements tend to have larger positive $\varepsilon_C^M$ values: As (13.0), Cu (4.1), Sn (9.4) (J.S.P.S., 1988), which are qualitatively consistent with the experimental results of Chabot et al. (2006). Because most siderophile elements are less compatible in liquid metal in the presence of C, W and Mo will respond in the opposite direction to that of most other elements if significant C is sequestered into core-forming metal (Jana and Walker, 1997a).

## 5. Discussion

*5.1. Accretion/core formation model prediction of BSE W and Mo concentrations*

The formation of Earth's core was a multi-stage heterogeneous process, during which accretion and differentiation occurred simultaneously. Modelling core formation as a heterogeneous process can rigorously consider the full range of accretionary processes, including, for example, the consequences of late accreted material mixed in from the outer Solar system. Unlike simple geochemical models of homogeneous core formation, which treat core-mantle equilibration as a single event, this requires a sophisticated modelling approach. The effects of impacts, changing impactor origins, and core formation throughout accretion must be considered in order to properly interpret BSE Mo and W concentrations.

We use the planetary accretion and differentiation model of Rubie et al. (2015) with the addition of S and the highly siderophile elements (HSEs) (Rubie et al., 2016) to investigate the behaviour of Mo and W in Earth during combined accretion and core formation. Our new expressions describing Mo and W partitioning behaviour are incorporated into the model, including the activity model for the metal. We also add C to the model of Rubie et al. (2016) assuming a constant C metal–silicate partition coefficient $D_C$ = 1000 (Hirschmann, 2016). $D_C$ has been shown experimentally to decrease with increasing pressure and temperature (Li et al. 2016; Malavergne et al., 2019; Fischer et al., 2020), as well as with increasing oxygen fugacity (Hirschmann, 2016; Malavergne et al., 2019). However, we found no changes to our results when we varied $D_C$ between 500 and 3000, which corresponds to more oxidised and reduced metal–silicate equilibration conditions, respectively (Hirschmann, 2016).

To briefly summarize the planetary accretion and differentiation model, the evolving composition of Earth's mantle and core were simulated by combining the dynamics of planetary accretion with the evolving chemistry of the cores and silicate portions of the impactors (for an in-depth description of the model see Rubie et al. 2015; 2016). The N-body accretion model begins with an initial distribution of planetary embryos and planetesimals in the protoplanetary disk and tracks their dynamics, gravitational interactions, and growth via collisions. We simulated "Grand Tack" terrestrial planet formation scenarios, whereby the inward-then-outward migration of Jupiter truncates the planetesimal disk which then results in a very successful reproduction of the mass and orbital properties of the terrestrial planets (Walsh et al., 2011; O'Brien et al., 2014; Jacobson and Morbidelli, 2014). Each collision delivers mass and energy to the growing planets, causing melting, magma ocean formation, and a discrete episode of core formation. After each impact, the projectile's mantle is mixed fully with the target's mantle, but a critical feature of the model is that the projectiles' cores only equilibrate with a small fraction of the target's mantle as they sink through a magma ocean, the volume of which is determined by the hydrodynamic model of Deguen et al. (2011; see also Rubie et al., 2015, Fig. 3). We assume that the metallic core of the projectile is fully emulsified as it sinks through the magma ocean (Deguen et al., 2014; Kendall and Melosh, 2016), so that the fraction of accreted metal that equilibrates with the small fraction of the silicate mantle is 100% (for a discussion of this assumption see Rubie et al., 2015). Following Rubie et al. (2015, 2016), a combination of rigorous mass balance and partitioning expressions is used to calculate the compositions of the metal and silicate portions of the planetary bodies after each equilibration event (Rubie et al., 2011).

Six Earth-like planets were selected from a suite of Grand Tack scenario terrestrial planet formation simulations (Jacobson and Morbidelli, 2014). These are the same six simulations presented in Rubie et al., 2015: (1) 4:1-0.25-7, (2) 4:1-0.5-8, (3) 8:1-0.25-2, (4) 8:1-0.8-8, (5) i-4:1-0.8-4, and (6) i-4:1-0.8-6; this paper refers to each simulation ordinally as 1-6. Each simulated Earth-like planet ends with a final mass and orbital distance from the Sun similar to that of Earth. From the N-body simulations, the entire growth history of each Earth-like planet is known including the accretion location of all the initial planetesimals and planetary embryos ultimately incorporated into the planet. All these initial bodies contain non-volatile elements (Si, Mg, Fe, Ni, Co, and Cr) in solar system (CI chondrite) relative abundances except for the most refractory elements which condense into solids at temperatures above the condensation temperatures of the major silicate phases and metal (Ca, Al, V, Nb, and Ta). Their concentrations are increased by 22% to match Earth's excess relative to CI chondrites (Palme & O'Neill 2014; this degree of enrichment is fully justified in Rubie et al., 2011).

There is a heliocentric oxidation gradient in the disk, which mimics what is seen amongst the chondritic meteorites and is necessary to obtain the correct mantle composition of Earth (Rubie et al., 2015). The oxidation gradient is such that the most reduced and volatile-depleted materials originate closest to the Sun, whereas the most oxidised and volatile-rich materials originate in the outer Solar system. The four parameters, which define the gradient and therefore the iron-rich metallic fraction of the starting bodies, are refined using a least squares regression, which fits the final abundances of the Earth-like planet's non-volatile lithophile and moderately siderophile elements (Si, Mg, Fe, Ni, Co, Cr, V, Nb, and Ta; see Rubie et al., 2015 for more details) to estimates of Earth's primitive mantle composition (Palme & O'Neill, 2014; see supplementary Figs. S4 and S5 and supplementary table S5 for best fit mantle abundances for each Earth-like planet). Despite the independent fitting of all six simulations, each best fit gradient has all of the Fe and about 10% of the Si appearing as metal interior to about 1.2 AU, an oxidation state comparable to that of enstatite chondrites (see Fig. 8 and supplementary Fig. S6). In all six accretion simulations, about 70% of the mass of material accreting to form the Earth-like planet originates at heliocentric distances of less than 1.2 AU (see Rubie et al. 2015, Fig. 1). Furthermore, fully oxidized material originates from the giant planet forming region (exterior to 4.5 AU in the model) and contributes less than 1% of each Earth-like planet's final mass. These fully oxidized outer solar system bodies contain no metallic core, but they do contain 20 wt.% water, consistent with carbonaceous chondrites and their C-type asteroid progenitors (Young, 2001): this results in a mantle water content of about 0.2 wt.% in each Earth-like planet, see supplemental Fig. S4.

The pressure of metal–silicate equilibration is also a fitted free parameter and is found to be about 70% of the target's core-mantle boundary pressure at the time of each accretional impact (Rubie et al., 2015), which is broadly consistent with calculations of impact-induced melting (de Vries et al., 2016). The temperature of metal–silicate equilibration is fixed to lie approximately midway between the peridotite liquidus and solidus at the equilibration pressure. The least squares refinement of the five free parameters (four defining the oxidation gradient and one defining the metal–silicate equilibration pressure) using nine elemental BSE abundance constraints results in excellent fits, as shown in supplementary Figs. S4 and S5. The fitted parameters are very similar to those of Rubie et al. (2015). Note that we did not include the effect of interaction with C on any element other than W and Mo. The changes in the abundances of the elements in supplementary Figs. S4-S6 between low- and high-C simulations are not due directly to the change in C abundance, but indirectly through the change in the fitted disk parameters such as the heliocentric gradient of Fe-metal abundance, which are different if attempting to match the low or high C mantle abundances.

In a separate second least squares minimisation, following the approach of Rubie et al. (2016), we model the BSE abundances of S and the highly siderophile elements (Pt, Pd, Ru, and Ir; Palme and O'Neill, 2014), but unlike in Rubie et al. (2016), we do not fit the concentration of S, nor any other element, in the simulated core; in other words, we make no assumption about core composition. In addition, we add Mo, W, and C to the model (Palme and O'Neill, 2014; Marty, 2012; Halliday, 2013). Mo and W were added by incorporating the metal–silicate partitioning models for Mo and W developed here (including the interaction parameters of Table 3). As described by Rubie et al. (2011, 2015), the most refractory non-volatile elements in starting bodies are enriched by a factor of 22% relative to CI chondrites. W and Mo are therefore also enriched by a factor of 22%. Since Mo and W are refractory, they are not explicitly associated with any free model fitting parameters. Thus, using eight estimated mantle abundances (Pt, Pd, Ru, Ir, S, Mo, W, and C) as constraints, we fit four free parameters (S and C heliocentric abundance gradients, pressure of silicate-sulphide equilibration, lifetime of post-giant impact mantle magma ocean), each of which are described in detail below. The final highly siderophile element abundances for each simulation are shown in supplementary Fig. S6 and Table S6.

Following Rubie et al. (2016), we assume that volatile elements condensed to an increasing extent as temperatures decreased in the protoplanetary disk – i.e. with increasing heliocentric distance. Thus, the concentration of S within the initial planetesimals and embryos is modelled as a simple linear gradient through the protoplanetary disk with an abundance of 5.35 wt.% (consistent with CI chondrites) in the outer solar system (at ≥ 4.5 AU). The distance in the inner solar system at which the S concentration linearly drops to zero from a value of 5.35 wt% at 4.5 AU is adjusted as a free parameter so that the final mantle S abundance of the Earth-like planet matches estimates of Earth's mantle abundance. The resulting fitted gradients (Fig. 8) predict that the highly-reduced material that originates at <1 AU contains low concentrations of S (according to one of our accretion simulations the S concentration at <1 AU is zero). In contrast, enstatite chondrites (ECs), which have been proposed to be Earth's main building blocks, are highly reduced with high S concentrations of up to 5–6 wt%. However,  et al. (2020) have shown that the ECs formed with relatively high volatile concentrations from *residual gas* in the solar nebular and are quite distinct from Earth's main building blocks which were highly reduced and volatile-poor. It has also been proposed that Mercury's highly reduced bulk composition includes a high S concentration, but this is currently very uncertain.

In addition to S, C is a necessary addition to the Rubie et al. (2016) model because it increases the siderophility of both W and Mo when C is dissolved in the metal liquid. Similar to S, the concentration of C within the initial planetesimals and embryos is modelled as a simple linear gradient through the protoplanetary disk with an abundance of 3.48 wt.% (consistent with CI chondrites) in the outer solar system (at ≥ 4.5 AU), and the distance at which the C concentration linearly drops to zero from a value of 3.48 wt% at 4.5 AU is adjusted as a free parameter to match the bulk silicate Earth abundance of C. However, there are two distinct groups of estimates of the bulk mantle or BSE C abundance: low (estimates range from 66 to 164 ppm; Hirschmann & Dasgupta, 2009; Halliday, 2013; Rosenthal et al., 2015; Hirschmann, 2016, 2018) and high (786 ± 308 ppm; Marty, 2012). The best fit S and C concentration gradients for each simulation 1-6 are shown in Figs. 8 and 9 for a high (786 ± 308 ppm; Halliday, 2013) and a low (66 ± 21 ppm) C mantle abundance esimate, respectively.

The partitioning of S between metal and silicate during core formation is modelled using Eq. 11 of Boujibar et al. (2014) (see Supplementary Information). Furthermore, a pervasive iron sulphide segregation event (the "Hadean Matte"; O'Neill, 1991) occurs during the cooling and crystallization of the global mantle magma ocean generated after each giant impact (i.e. the multistage scenario of Rubie et al. 2016). This results in a new core addition rich in highly siderophile elements, because they partition strongly into the exsolved iron sulphide at the lower pressures and temperatures of mantle magma ocean crystalization. Modelling these events requires the fitting of two additional free parameters: the effective pressure of silicate-sulfide equilibration (the equilibration temperature is assumed to lie approximately midway between the peridotite liquidus and solidus at the equilibration pressure) and the time duration of the mantle magma ocean prior to solidification, which is important because planetesimal impacts after crystallisation of the final deep mantle magma ocean are treated as a late veneer that does not contribute to core formation. The duration of the final magma ocean is refined to be about 25 Myr but there was considerable dispersion set by the stochastic history of late accretion on each Earth-like planet.

The refined equilibration pressure of exsolved sulphide droplets with the magma ocean is determined to be much lower than that of metal–silicate equilibration. In the case of metal–silicate equilibration, this is assumed to occur between accreted metal and a plume of entrained silicate liquid at the base of a magma ocean (Deguen et al., 2011; Rubie et al., 2015, their Fig. 3) – which could potentially be at or close to the core-mantle boundary in the case of giant impacts (de Vries et al., 2016). Sulphide segregation is quite different. FeS exsolves during the cooling and crystallization of a magma ocean which causes it to become saturated in FeS. This occurs because the sulfur concentration at sulfide saturation (SCSS) decreases strongly with decreasing temperature (e.g. Mavrogenes and O'Neill, 1999; Liu et al., 2007; Laurenz et al., 2016). Thus, exsolution of FeS droplets is a pervasive process that occurs over a large depth interval. To understand the chemical consequences of equilibration between these dispersed FeS droplets as they sink to the core and the convecting magma ocean is clearly a complex problem. Rubie et al. (2016) took a highly-simplified approach by assuming that equilibration could be modelled at a single effective pressure which is constrained by the final mantle concentrations of the HSEs and S. This pressure is considerably lower than the metal–silicate equilibration pressure and is found to be about 30% of the target's core-mantle boundary pressure.

We added Mo and W to the model of Rubie et al. (2016) as before. In all simulations, we obtained good matches to the BSE abundances of W, Mo, and C when fitting the high C abundance estimate for the BSE, but we obtained much poorer fits for W and C when attempting to fit the low C abundance BSE estimate (Fig. 10). S is better matched when fitting the low C BSE estimate than when fitting the higher value, but is generally within two-sigma of the S BSE abundance estimate of Palme and O'Neill (2014) and below the highest estimates (e.g. Sun 1982). From these results, it is clear that the importance of the effects of C and S on the partitioning of W and Mo cannot be overstated. If C is not present in the equilibrating liquids, then too much W will remain in the silicate liquid after equilibration and it will accumulate too much in the mantle over time (Figs. 11 and 12). This is what occurs when the C abundance is suppressed by varying the zero-abundance distance of the C gradient in the disk so that C is brought nearly exclusively from outer disk planetesimals. When the C abundance in the disk is increased, thereby increasing the abundance in Earth's mantle to be consistent with the high (786 ppm) estimate, the abundance of W in the mantle decreases to match the best estimates of Earth's silicate W abundance. It's important to note that the effect of C is regulated by the presence of S, which has the opposite effect of C on the partitioning of W. Thus, if S accumulates too quickly relative to C in an Earth-like planet's mantle during accretion, then the abundance of W becomes too great relative to Earth's BSE value. The accretion of S must remain sufficient enough though, so that S will be exsolved during magma ocean crystallization in order to explain the abundances of the HSEs (Rubie et al. 2016). Thus, in order to match the abundance of Mo, S, and HSEs in the BSE, the final C mantle abundance must always be greater than 300 ppm and to match the abundance of W, the final mantle C abundance must be greater than 700 ppm.

### 5.2. Implications for magma ocean crystallisation and fractionation processes in the case of a low C mantle abundance

If Earth's mantle abundance of C is low (Hirschmann & Dasgupta, 2009; Halliday, 2013; Rosenthal et al., 2015; Hirschmann, 2016, 2018), then the model predicts a BSE in which W and C are overabundant when all other elements considered have Earth-like concentrations. For W, the simplest explanation is that the BSE W content is not well-enough known and the values inferred from element ratios in basaltic melts (Arevalo and McDonough, 2008; König et al., 2011) are significantly underestimated. For example, Rizo et al. (2017) presented compositions of fertile mantle xenoliths that contain considerably higher W concentrations than published BSE values, although the highest concentrations may be attributable to metasomatic enrichment. Large uncertainties also exist for the BSE Mo concentration, with different terrestrial sample types and questions around melting and refertilisation behaviour resulting in both lower (23 ppb, Greber et al., 2015) and higher (113 – 180 ppb; Liang et al., 2017) suggested BSE values than the 47 ppb of Palme and O'Neill (2014). The large 2σ uncertainty envelope of the Palme and O'Neill estimate encompasses most of this range. However, we find that even with the highest estimates of the C metal–silicate partition coefficient $D_C$ = 3000 (Hirschmann, 2016), C is significantly over-abundant in the mantle.

There are two main reasons for high mantle concentrations of W and C according to the model results. (1) When these elements are delivered in fully oxidized bodies that originate from beyond 4.5 AU, there is no core forming event that segregates them to the core because such bodies contain no metal. Thus W and C remain in the target's mantle. (2) When metal-bearing bodies impact Earth, the metal only equilibrates with a small fraction of the magma ocean/mantle before segregating to the core. Metal–silicate equilibration is thus very inefficient at removing the elements to the core. This raises the question: would the modelled BSE W and C concentrations still be anomalously high if the extent of Earth's mantle participating in metal–silicate equilibration were greater? We investigate this by running the model with different extents of mantle equilibration, and found that this does not solve the issue (Fig. 13). According to the Deguen et al. (2011) model, core forming liquids from giant impacts should on average equilibrate with about 7% of the target mantle. Enhancing this fraction does indeed decrease the final mantle concentrations of W concentrations because more W can be leached out of the terrestrial mantle during each core formation event. However, even at very high equilibrating mantle fractions, the model W abundance is about a factor of two greater than observed. In addition, concentrations of non-volatile elements deviate significantly from BSE values as indicated by an increase in the reduced chi squared (see also Rubie et al., 2015). Thus, a small fraction of equilibration most realistically reproduces Earth, in agreement with laboratory experiments (Deguen et al. 2011, Landeau et al. 2017), and forces us to reject the hypothesis that the mismatch between model W concentrations and BSE W concentrations is an artefact of choosing a low extent of equilibration. This leaves a significant problem regarding W in the mantle.

Here, we speculate on possible subsequent processing that is not included in the model. In other words, the model outputs are correct and the early silicate Earth really was overabundant in W and C relative to present day concentrations. In this case, some process must have later removed this excess. The cause of the overabundance is primarily the localised and limited extent of equilibration between the cores of late impactors with the mantle magma ocean, resulting in less removal of silicate W and C to the core than would be expected for the bulk Earth's $fO_2$. A pervasive late-stage metal segregation would be one way to rectify this. Metal precipitating throughout Earth's molten or crystallising mantle would strip it of excess siderophile elements and sequester them to the core. Small metal droplets would fully equilibrate with the silicate at the prevalent oxygen fugacity, removing W, C, and other siderophile elements. The efficiency of a pervasive late-stage metal in removing W would be increased if it were C-rich, for example, in areas of localised high C concentrations. Moreover, added C would diminish the ability of the metallic liquid to sequester most other siderophile elements (relative to a C-free liquid). The saturation of a dispersed carbide ($Fe_7C_3$) phase would have a similar effect if we assume that partitioning between silicate liquid and carbide is similar to that between silicate liquid and the stoichiometrically-similar carbon-saturated metal. This is in contrast to a high metallic S content, which would have the opposite effect on the metal–silicate partitioning of W (Wood et al., 2014), so a sulphide "Hadean Matte" cannot be responsible (Rubie et al., 2016).

The precipitation of metallic iron is an expected consequence of deep magma ocean formation and crystallisation. Frost et al. (2004) demonstrated through experiments that aluminium-bearing bridgmanite has such a strong capacity for $Fe^{3+}$ that its crystallisation will drive the disproportionation of $Fe^{2+}$ to $Fe^{3+}$ plus metallic Fe, even under very reduced conditions. They calculated that the crystallisation of the lower mantle must have driven the precipitation of around 1 wt.% metallic iron. Such a disproportionation reaction likely occurs in the silicate melt itself, as the ferric iron component is stabilised by high pressures (Armstrong et al., 2019). The formation, cooling and crystallisation of the magma ocean(s) could therefore drive the exsolution of dispersed metal, stripping the mantle of its excess W. This metal would be liquid at the relevant *P-T* conditions, so would percolate to the core. We calculate that the precipitation of 0.5 wt.% of pure Fe metal through disproportionation could remove 56% of W and 44% of Mo from the mantle in a simplistic heterogeneous accretion scenario (see supplementary information). Along with W, other siderophile elements such as C could similarly be sequestered.

The accretion, differentiation and potential iron disproportionation scenario presented here has interesting implications for planetary accretion and early differentiation. Our results have implications for the interpretation of isotope data, and likewise must fit constraints imposed by that data. If the BSE was indeed initially higher in W and C than today, then these results demonstrate the need to consider the effects of late fractionation processes on the composition of the BSE (which is known to be important for the HSE elements; Laurenz et al., 2016; Rubie et al., 2016).

The model presented here involves a number of simplifying assumptions. For example, it is assumed that volatile element concentrations in starting bodies are a linear function of heleocentric distance, which may not be realistic. It is also assumed that all metal-bearing starting bodies underwent core-mantle differentiation prior to being accreted to protoplanets. In reality some bodies must have been undifferentiated and would have released small dispersed metal grains into the magma ocean; the effect of this on the evolving compositions of the simulated planets remains to be determined. In future, incorporating isotopic constraints (including the Hf-W isotopic system) could be used to reduce the number of simplifying assumptions.

### 5.3. Implications for isotope systematics

Late core formation has important implications for the interpretation for the short-lived $^{182}$Hf–$^{182}$W system. An increased initial W abundance would decrease the initial BSE Hf/W and so radiogenic $^{182}$W would accumulate in the mantle relative to non-radiogenic $^{184}$W at a correspondingly slower rate. Note that the $^{182}$W/$^{184}$W ratio would not be disturbed by dispersed metal saturation if that saturation occurred after most of the short-lived $^{182}$Hf had decayed (~50 Myr). Elevated $^{182}$W/$^{184}$W anomalies would be expected to be smaller, and ratios would only be resolvable from a shorter timescale than currently thought: in Earth, this may aid the reconciliation of the contrasting $^{182}$Hf–$^{182}$W and $^{146}$Sm–$^{142}$Nd records (Kleine et al., 2009), restricting the apparent timing of core-formation to a narrower time period after Earth's formation. In addition, the W content is increased by partial equilibration with impactors, which is compatible with evidence of preserved $^{182}$W heterogeneities in the mantle (Touboul et al., 2012). Zube et al. (2019) found that the Grand Tack model creates Earth-like planets too quickly so that too much $^{182}$W remains in their mantles today to match Earth. However, they assumed a constant partition coefficient of W. Here, and in previous experimental studies, it has been shown that the partitioning of W depends predominantly on $fO_2$ and additionally on temperature and on the presence of C and S in the equilibrating liquids. This complicates the analysis and demands further study. Wade and Wood (2016) incorporated $^{182}$Hf–$^{182}$W isotopic constraints from the Earth-Moon system in an accretion and differentiation model, and suggested that such constraints were best met with a reduced moon-forming impactor, in contrast to our conclusions. The implications for the W-Hf system, core formation ages and the moon-forming impactor are clearly not yet resolved.

We predict that the majority of W and Mo present in Earth's mantle, along with S, was delivered relatively late during its accretion, by a higher fraction of more oxidised impactor material. Heterogeneity exists across many isotopic systems in the meteorite record, with an apparent dichotomy between carbonaceous and non-carbonaceous chondrite compositions (Kruijer et al., 2020). Given that carbonaceous chondrites originate further from the sun, the isotopic identity of the BSE, like its elemental composition, can also record the timing of mixing and accretion between materials originating from different locations within the Solar System (Dauphas, 2017). Various isotopic systems, including Mo isotopes, may qualitatively agree with our model in that they imply the later addition of more carbonaceous chondrites relative to non-carbonaceous chondrites (Budde et al., 2019; Kruijer et al., 2020).

An important future line of enquiry would thus be to resolve these different records, to reconcile partitioning behaviour and accretionary events with the elemental and isotopic identity of Earth's building blocks. Our interpretation therefore requires testing in the future in order to 1) check its effect on the concentrations of other moderately siderophile elements in the mantle (the effect of C interaction and disproportionation on them has not been considered here), and 2) to understand in detail its implications for, and compatibility with, $^{182}$Hf–$^{182}$W systematics (and potentially other isotope systems).

## 6. Conclusions

We have re-evaluated the metal–silicate partition coefficients $D_{Mo}$ and $D_W$ and the $fO_2$-independent exchange coefficients $K_D^{Mo}$ and $K_D^W$ based on 48 new experiments and a large compiled dataset of previously-published results. The control of $fO_2$ on $D_{Mo}$ and $D_W$ can be predicted if the valence of Mo and W in silicate melts is well-constrained: our results confirm that $Mo^{4+}$ and $W^{6+}$ are the stable cations over the full experimental $fO_2$ range.

$K_D^{Mo}$ is sensitive to pressure and temperature, but these two intensive parameters mostly counteract one another at experimentally-attained conditions such that $K_D^{Mo}$ decreases only gradually with increasing depth in Earth. Our parametrisation of $K_D^{Mo}$ gives fairly similar results to all previous studies, so changes in $K_D^{Mo}$ with pressure and temperature seem known with reasonable certainty. From our compilation of metal–silicate partitioning data, it is not possible to distinguish correlations between $K_D^{Mo}$ and silicate melt composition within uncertainty, so we make an approximation that $K_D^{Mo}$ is insensitive to silicate melt composition. $K_D^{Mo}$ was, however, found to depend strongly on the light element content of the metallic phase.

$K_D^W$ is sensitive to temperature but not pressure, becoming more siderophile with depth in Earth's mantle because of the temperature increase; this conclusion is broadly in agreement with some other studies (Siebert et al., 2011; Wade et al., 2012). However, different studies have more divergent models for $K_D^W$ and its changes with $P$ and $T$ (see supplementary fig. S3), perhaps because of the higher sensitivity of $K_D^W$ to both silicate and metal melt composition compared to $K_D^{Mo}$, making W behaviour somewhat more challenging to characterise than Mo. If W experimental data are restricted to those with silicate compositions similar to magma ocean compositions (broadly pyrolite-like), compositional effects on partitioning become weaker and can be neglected to some extent. Note that this excludes the use of data based on basaltic compositions. $K_D^W$ was found to be particularly sensitive to the silicate melt CaO content, in agreement with previous suggestions of possible complexing behaviour or melt polymerisation control on the activity coefficient of $WO_3$ (O'Neill et al., 2008; Steenstra et al., 2017).

$K_D^W$ and $K_D^{Mo}$ are both sensitive to the carbon content of the metal phase, such that W and Mo become increasingly siderophile when the metal is carbon-bearing. However, this effect is less dramatic than is predicted by interpolating linearly between carbon-free and carbon-saturated partitioning experiments (cf. Cottrell et al., 2009). Our newly-fitted epsilon interaction parameters can be used to predict the effect of C on Mo and W partitioning up to around 3 wt. % C in the metal.

We added the updated parametrisations of partitioning coefficients of W and Mo as well as the interaction parameters of W and Mo with C to the combined accretion and core-formation model of Rubie et al. (2015, 2016), in order to interpret the concentrations of Mo and W in the BSE in terms of planetary formation processes. In this model, oxidation and volatility gradients extend through the solar nebular. In this protoplanetary disk, Earth accretes heterogeneously and undergoes a multi-stage core formation process. We found that C plays an important role in the partitioning of W and Mo into the core, in agreement with previous suggestions (Jana and Walker, 1997a; Righter and Drake, 1999; Chabot et al., 2006; Cottrell et al., 2009). The model results show that W and Mo concentrations require the early accreting Earth to be sulfur-depleted and carbon-enriched so that W and Mo are efficiently partitioned into Earth's core and do not accumulate in the mantle. Indeed, without considering the presence of C, too much Mo and W accumulates in the mantle to match terrestrial abundances. In order for C to have a substantial enough effect on the partitioning of Mo and W to bring their final abundances down into agreement with measured values, the final C abundance is more consistent with the high C BSE abundance of about 800 ppm (Marty, 2012) than the low C abundance estimates ranging from 70 to 140 ppm (Hirschmann & Dasgupta, 2009; Halliday, 2013; Hirschmann, 2018). Alternatively, Earth's mantle C abundance was very high directly after planetary accretion, but then a subsequent core formation event triggered by disproportionation of $Fe^{2+}$ to $Fe^{3+}$ plus metallic Fe during deep magma ocean formation and crystallization may sequester more W and C from the mantle into the core and leave behind abundances that are consistent with measurements that favour the lower abundance of C. These concepts are required to reconcile model results with the BSE composition, and should be testable by applying constraints from Hf-W isotope systematics.

## 7. Acknowledgements

We would like to thank Andreas Audétat for performing the LA-ICP-MS measurements of the quenched silicate melts, Tiziana Boffa Ballaran for performing the XRD analysis of the carbide standard, Detlef Krauße for assistance with the electron microprobe, Hubert Schulze for sample preparation, Jon Wade for providing the data of Walter and Thibault (1995), and Alessandro Morbidelli for discussions. We also thank Nicolas Dauphas and Wim van Westrenen for editorial handling and additional perceptive comments, and four anonymous reviewers and Jon Wade for their thorough and constuctive reviews that enabled us to significantly increase the quality of our work. We acknowledge financial support from the European Research Council (ERC) Advanced Grant "ACCRETE" (Contract No. 290568) and the German Science Foundation (DFG) Priority Programme SPP1833 "Building a Habitable Earth" (RU 1323/10-1) and SPP1385 "The First 10 Million Years of the Solar System – a Planetary Materials Approach" (RU 1323/2).

## Tables

Table 1: Experimental run conditions.

| Group | Experiment | P GPa | T K | Duration min |
|---|---|---|---|---|
| AV | H3371a | 11 | 2773 | 2.0 |
| AV | H3371b | 11 | 2773 | 2.0 |
| AV | H3372a | 11 | 2478 | 2.3 |
| AV | H3372b | 11 | 2478 | 2.3 |
| AV | H3586a | 11 | 2609 | 2.2 |
| AV | H3586b | 11 | 2609 | 2.2 |
| AV | Z1009b | 11 | 2619 | 2.0 |
| AV | Z1011b | 11 | 2605 | 2.0 |
| AV | Z822b | 11 | 2573 | 3.0 |
| AV | Z859a | 18 | 2717 | 2.1 |
| AV | Z865a | 18 | 2834 | 1.0 |
| AV | Z916b | 11 | 2580 | 2.0 |
| AV | Z929a | 20 | 2799 | 1.5 |
| AV | Z950a | 21 | 2904 | 1.5 |
| AV | Z970a | 21 | 2911 | 1.5 |
| AV | Z970b | 21 | 2911 | 1.5 |
| ESJ | A962b** | 1.5 | 1923 | 30.0 |
| ESJ | A961a-L | 1.5 | 1923 | 30.0 |
| ESJ | A961a-R | 1.5 | 1923 | 30.0 |
| ESJ | A961b-L | 1.5 | 1923 | 30.0 |
| ESJ | A961b-R | 1.5 | 1923 | 30.0 |
| LRW | A81 | 3.14 | 2223 | 30 |
| LRW | A73 | 3.14 | 2223 | 30 |
| LRW | A74 | 3.14 | 2223 | 30 |
| LRW | H2603 | 10 | 2693 | 1.5 |
| LRW | V378 | 10 | 2693 | 1.5 |
| LRW | H2617 | 15 | 2573 | 1.5 |
| LRW | H2775 | 6 | 2273 | 2 |
| LRW | H2735 | 10 | 2693 | 1.5 |
| LRW | H2748 | 12 | 2693 | 1.5 |
| LRW | H2749 | 15 | 2693 | 1.5 |
| LRW | H2755 | 19 | 2593 | 1.5 |
| LRW | A129 | 3.14 | 2223 | 30 |
| LRW | 4A*** | 3.14 | 2223 | 30 |
| LRW | 4At2*** | 3.14 | 2223 | 15 |
| LRW | 4At3*** | 3.14 | 2223 | 45 |
| LRW | 4At4*** | 3.14 | 2223 | 71 |
| LRW | 4At5*** | 3.14 | 2223 | 100 |
| YN | S5700 | 25 | 2960 | 3 |
| YN | S5703 | 25 | 2970 | 3 |
| YN | S5706 | 15 | 2810 | 3 |
| YN | H3346 | 25 | 2870 | 3 |
| YN* | S4818 | 25 | 2850 | 4 |
| YN* | S4842 | 25 | 2950 | 3 |
| YN* | S4858Fo95-A | 25 | 2950 | 2 |
| YN* | S4858Fo70-B | 25 | 2950 | 2 |
| YN* | S4933 | 25 | 2950 | 2 |
| YN | S5056-Spoor | 25 | 2870 | 3 |
| YN* | S5069 | 15 | 2720 | 3 |
| YN* | S5134 | 15 | 2802 | 3 |
| YN* | S5139 | 15 | 2807 | 3 |
| YN* | S5140 | 15 | 2807 | 3 |
| YN* | S5157 | 20.5 | 2727 | 2.5 |

*reported by Fischer et al. (2015), trace elements in silicate re-analysed by ICPMS
**graphite capsule (all other experiments performed in MgO capsules)
***time series, S-bearing starting material

Table 2a, b: Major and trace element analyses of experimentally quenched phases, along the number of analytical points. LA-ICPMS analyses are indicated by italics; remaining data is EPMA. a) Silicate, b) metal, including time-series experiments. In most experiments there are additional trace elements present and analysed which are not shown for clarity – the complete analyses including these trace elements are provided in Supplementary Table S2.

| Group | Experiment | n EPMA | Na₂O | 1σ | MgO | 1σ | Al₂O₃ | 1σ | SiO₂ | 1σ | CaO | 1σ | FeO | 1σ | SO₂ | 1σ | Total | n ICPMS | Mo | 1σ | W | 1σ | other |
|---|---|---|---|---|---|---|---|---|---|---|---|---|---|---|---|---|---|---|---|---|---|---|---|
| | | | Silicate phase, wt. % | | | | | | | | | | | | | | | | Silicate phase, ppm | | | | |
| AV | H3371a | 34 | | | 47.6 | 5.6 | 3.05 | 2.03 | 39.3 | 2.5 | 4.64 | 2.68 | 5.67 | 1.07 | *0.16* | *0.01* | 100.3 | 4 | *35.3* | *4.7* | *180* | *19* | *311* |
| AV | H3371b | 40 | | | 47.5 | 3.1 | 2.68 | 0.93 | 41.8 | 0.8 | 4.33 | 1.67 | 4.69 | 1.05 | | | 101.0 | 4 | *8.9* | *2.3* | *49* | *15* | *257* |
| AV | H3372a | 12 | | | 45.4 | 4.1 | 3.23 | 1.39 | 41.5 | 0.6 | 3.84 | 1.74 | 7.07 | 1.36 | *0.12* | *0.02* | 101.1 | 5 | *27.7* | *3.7* | *176* | *11* | *308* |
| AV | H3372b | 8 | | | 45.5 | 8.4 | 3.52 | 2.84 | 41.9 | 0.3 | 4.41 | 3.67 | 6.14 | 1.71 | | | 101.5 | 5 | *18.2* | *1.4* | *50* | *9* | *220* |
| AV | H3586a | 32 | | | 48.2 | 2.7 | 2.55 | 1.05 | 39.5 | 0.8 | 4.08 | 1.58 | 5.37 | 1.24 | | | 99.6 | 6 | *4.1* | *1.8* | *61* | *16* | *142* |
| AV | H3586b | 28 | | | 45.9 | 4.0 | 4.12 | 1.73 | 39.1 | 1.3 | 6.52 | 2.47 | 4.57 | 0.83 | | | 100.1 | 4 | *3.4* | *1.0* | *34* | *10* | *101* |
| AV | Z1009b | 10 | | | 47.4 | 1.6 | 3.20 | 0.66 | 39.5 | 0.4 | 3.09 | 0.92 | 5.96 | 0.77 | *0.20* | *0.03* | 99.3 | 6 | *20.2* | *9.5* | *273* | *86* | *242* |
| AV | Z1011b | 10 | | | 45.1 | 7.0 | 4.07 | 2.51 | 40.3 | 0.3 | 4.68 | 3.26 | 5.30 | 1.54 | | | 99.5 | 6 | *2.9* | *1.2* | *18* | *6* | *121* |
| AV | Z822b | 30 | | | 45.2 | 6.3 | 3.49 | 2.13 | 42.2 | 0.5 | 4.30 | 2.75 | 5.72 | 1.18 | | | 100.9 | 5 | *3.4* | *0.6* | *27* | *3* | *153* |
| AV | Z859a | 30 | | | 40.0 | 3.8 | 4.40 | 1.06 | 44.2 | 1.1 | 4.34 | 1.11 | 6.98 | 0.71 | | | 99.9 | 5 | *3.6* | *0.3* | *27* | *6* | *203* |
| AV | Z865a | 28 | | | 41.1 | 2.8 | 3.71 | 0.33 | 43.9 | 1.3 | 4.01 | 0.90 | 6.32 | 0.48 | | | 99.1 | 5 | *4.7* | *1.7* | *31* | *15* | *186* |
| AV | Z916b | 15 | | | 47.5 | 2.9 | 3.01 | 1.12 | 41.4 | 0.4 | 3.11 | 1.13 | 5.51 | 0.72 | | | 100.5 | 4 | *6.9* | *4.7* | *62* | *28* | *203* |
| AV | Z929a | 28 | | | 40.7 | 4.3 | 4.33 | 1.22 | 42.7 | 1.7 | 4.00 | 1.39 | 6.69 | 0.60 | | | 98.5 | 5 | *5.1* | *1.1* | *43* | *5* | *217* |
| AV | Z950a | 9 | | | 37.9 | 2.2 | 3.83 | 0.16 | 44.7 | 1.8 | 4.26 | 1.02 | 7.37 | 0.65 | | | 98.0 | 5 | *10.3* | *2.4* | *43* | *10* | *348* |
| AV | Z970a | 20 | | | 39.3 | 1.8 | 4.27 | 0.37 | 44.2 | 1.2 | 4.58 | 0.65 | 6.82 | 0.59 | | | 99.2 | 5 | *7.1* | *1.0* | *29* | *4* | *261* |
| AV | Z970b | 20 | | | 39.1 | 2.1 | 4.29 | 0.47 | 45.0 | 1.4 | 4.45 | 0.70 | 6.39 | 0.50 | | | 99.2 | 5 | *5.5* | *1.1* | *22* | *5* | *206* |
| ESJ | A962b | 25 | | | 12.8 | 0.1 | 21.14 | 0.11 | 44.3 | 0.2 | 10.39 | 0.06 | 8.60 | 0.07 | | | 97.2 | 5 | *0.36* | *0.04* | *11.5* | *0.4* | |
| ESJ | A961a-L | 71 | | | 27.1 | 4.7 | 15.65 | 3.49 | 37.4 | 1.7 | 13.99 | 2.51 | 5.40 | 0.32 | | | 99.6 | 5 | *0.71* | *0.15* | *55.0* | *2.0* | |
| ESJ | A961a-R | 76 | | | 27.8 | 5.4 | 14.35 | 2.18 | 36.5 | 0.6 | 14.21 | 2.82 | 6.39 | 0.36 | | | 99.3 | 6 | *2.01* | *0.21* | *140.5* | *3.7* | |
| ESJ | A961b-L | 77 | | | 28.1 | 6.8 | 14.89 | 2.94 | 37.1 | 0.6 | 13.65 | 3.45 | 5.74 | 0.40 | | | 99.5 | 6 | *1.07* | *0.08* | *73.6* | *3.2* | |
| ESJ | A961b-R | 73 | | | 29.2 | 7.1 | 14.08 | 3.19 | 36.4 | 1.0 | 13.59 | 3.78 | 6.39 | 0.54 | | | 99.6 | 5 | *2.67* | *0.19* | *167.8* | *5.6* | |
| LRW | A81 | 17 | 0.55 | 0.12 | 30.2 | 2.7 | 8.84 | 1.08 | 39.8 | 0.5 | 9.55 | 1.33 | 9.48 | 0.42 | | | 98.5 | 5 | | | *144* | *10* | *2214* |
| LRW | A73 | 18 | 0.76 | 0.22 | 35.6 | 3.2 | 9.64 | 1.59 | 34.1 | 0.9 | 11.24 | 2.32 | 8.23 | 0.84 | 0.46 | 0.17 | 100.0 | 4 | | | *876* | *46* | *1217* |
| LRW | A74 | 10 | 0.59 | 0.09 | 31.0 | 1.0 | 9.96 | 0.67 | 34.7 | 0.2 | 13.13 | 0.92 | 11.00 | 0.41 | 0.47 | 0.18 | 100.8 | 6 | | | *1391* | *278* | *1687* |
| LRW | H2603 | 36 | 0.34 | 0.04 | 40.5 | 1.7 | 4.77 | 0.48 | 42.6 | 0.6 | 8.23 | 0.83 | 4.63 | 0.30 | 0.31 | 0.10 | 101.3 | 5 | | | *1769* | *145* | *2947* |
| LRW | V378 | 13 | 0.37 | 0.06 | 40.6 | 1.1 | 3.70 | 0.40 | 42.8 | 0.6 | 8.40 | 0.85 | 4.95 | 0.66 | 0.21 | 0.06 | 101.0 | 3 | | | *53* | *15* | *1886* |
| LRW | H2617 | 30 | 0.40 | 0.03 | 38.7 | 2.9 | 4.91 | 0.26 | 43.8 | 2.1 | 7.67 | 0.54 | 4.85 | 0.43 | 0.16 | 0.04 | 100.5 | 3 | | | *24* | *6* | *677* |
| LRW | H2775 | 24 | 0.23 | 0.06 | 38.5 | 3.1 | 8.42 | 2.37 | 37.5 | 1.5 | 8.20 | 1.83 | 6.82 | 0.83 | 0.31 | 0.13 | 100.0 | 6 | | | *22* | *7* | *1078* |

| Group | Sample | | | | | | | | | | | | | | | | | | | |
|---|---|---|---|---|---|---|---|---|---|---|---|---|---|---|---|---|---|---|---|---|
| LRW | H2735 | 11 | 0.20 | 0.08 | 42.6 | 4.2 | 3.69 | 1.30 | 40.7 | 1.3 | 5.69 | 2.17 | 8.06 | 1.64 | 0.21 | 0.20 | 101.2 | 5 | *318* | *15* | *1323* |
| LRW | H2748 | 11 | 0.20 | 0.06 | 45.6 | 2.5 | 2.76 | 0.83 | 42.4 | 0.6 | 5.34 | 1.62 | 4.36 | 0.37 | 0.15 | 0.06 | 100.8 | 3 | *46* | *9* | *876* |
| LRW | H2749 | 18 | 0.35 | 0.07 | 40.0 | 3.1 | 4.26 | 0.94 | 44.0 | 0.8 | 7.31 | 1.56 | 5.36 | 0.52 | 0.20 | 0.07 | 101.4 | 6 | *108* | *16* | *1230* |
| LRW | H2755 | 10 | 0.39 | 0.10 | 36.1 | 2.9 | 6.44 | 0.96 | 44.2 | 1.5 | 7.47 | 1.38 | 6.19 | 0.97 | 0.04 | 0.02 | 100.9 | 6 | *88* | *8* | *919* |
| LRW | A129 | 20 | 1.55 | 0.35 | 15.4 | 6.2 | 15.32 | 1.83 | 40.3 | 0.7 | 15.51 | 2.84 | 10.97 | 0.65 | 0.50 | 0.19 | 99.6 | 5 | *1167* | *121* | *2506* |
| LRW | 4A\*\*\* | 14 | 0.57 | 0.18 | 29.2 | 3.6 | 9.10 | 1.46 | 39.0 | 0.5 | 10.13 | 1.71 | 10.54 | 0.61 | 0.56 | 0.18 | 99.1 | 6 | *3387* | *736* | *1523* |
| LRW | 4At2\*\*\* | 15 | 0.72 | 0.83 | 25.8 | 6.9 | 11.65 | 2.87 | 39.1 | 0.9 | 11.83 | 3.28 | 10.9 | 1.82 | 0.51 | 0.35 | 100.6 | 7 | *6283* | *709* | *2072* |
| LRW | 4At3\*\*\* | 10 | 0.9 | 0.71 | 17.4 | 8.3 | 15.04 | 2.51 | 39.7 | 1.0 | 16.11 | 3.25 | 10.96 | 1.65 | 0.65 | 0.37 | 100.8 | 5 | *6170* | *333* | *2767* |
| LRW | 4At4\*\*\* | 27 | 1.19 | 0.5 | 23.7 | 9.1 | 13.31 | 4.57 | 34.3 | 2.3 | 16.18 | 4.99 | 11.14 | 1.57 | 0.6 | 0.33 | 100.4 | 4 | *8869* | *514* | *2655* |
| LRW | 4At5\*\*\* | 7 | 0.82 | 0.13 | 34.6 | 2.5 | 10.65 | 2.55 | 34.9 | 1.5 | 12.79 | 1.56 | 6.01 | 1.2 | 0.81 | 0.4 | 100.6 | 7 | *3777* | *61* | *1582* |
| YN | S5700 | 20 | | | 41.8 | 1.4 | | | 48.2 | 1.5 | | | 8.32 | 0.52 | 0.73 | 0.05 | 99.6 | | *261* | *218* | *9540* |
| YN | S5703 | 5 | | | 36.0 | 1.7 | | | 46.9 | 1.3 | | | 13.80 | 1.73 | 1.09 | 0.24 | 98.8 | | *380* | *326* | *15539* |
| YN | S5706 | 39 | | | 41.4 | 2.5 | | | 40.4 | 0.8 | | | 14.56 | 1.80 | 0.92 | 0.19 | 98.0 | | *414* | *376* | *12075* |
| YN | H3346 | 18 | | | 43.5 | 4.9 | 0.02 | 0.00 | 49.9 | 5.5 | | | 5.96 | 1.44 | 0.88 | 0.20 | 100.9 | 3 | *81* | *37* | *11665* |
| YN\* | S4818 | 10 | | | 41.7 | 1.8 | 0.03 | 0.00 | 52.9 | 1.7 | | | 5.06 | 0.41 | 0.48 | 0.03 | 100.7 | 3 | *122* | *75* | *7700* |
| YN\* | S4842 | 17 | | | 40.7 | 2.1 | 0.02 | 0.00 | 52.2 | 2.1 | | | 6.51 | 0.55 | 0.61 | 0.04 | 100.7 | 5 | *41* | *10* | *9752* |
| YN\* | S4858Fo95-A | 17 | | | 45.1 | 0.8 | 0.03 | 0.00 | 50.0 | 0.9 | | | 3.12 | 0.27 | 0.75 | 0.04 | 99.6 | 3 | *5.0* | *0.6* | *10352* |
| YN\* | S4858Fo70-B | 29 | | | 39.6 | 2.7 | 0.54 | 0.04 | 43.9 | 2.6 | | | 13.66 | 0.99 | 0.67 | 0.07 | 99.1 | 5 | *244* | *32* | *11074* |
| YN\* | S4933 | 34 | | | 43.6 | 1.1 | 0.68 | 0.00 | 49.4 | 1.2 | | | 6.69 | 0.44 | 0.45 | 0.02 | 101.3 | 2 | *20.5* | *1.5* | *6703* |
| YN | S5056-Spoor | 13 | | | 39.4 | 2.2 | 0.54 | 0.08 | 47.7 | 2.6 | | | 12.13 | 1.66 | 0.63 | 0.15 | 101.0 | 3 | *239* | *123* | *10036* |
| YN\* | S5069 | 19 | | | 48.7 | 2.3 | 0.03 | 0.01 | 46.4 | 1.3 | | | 5.13 | 0.63 | 0.43 | 0.08 | 101.2 | 1 | *20.2* | *3.7* | *6733* |
| YN\* | S5134 | 56 | | | 45.2 | 2.4 | 0.57 | 0.05 | 41.5 | 1.5 | | | 12.63 | 1.15 | 0.47 | 0.07 | 100.9 | 5 | *337* | *70* | *7953* |
| YN\* | S5139 | 25 | | | 51.4 | 2.2 | 0.03 | 0.00 | 46.3 | 1.6 | | | 1.77 | 0.22 | 0.47 | 0.05 | 100.3 | 5 | *2.8* | *0.6* | *6457* |
| YN\* | S5140 | 22 | | | 50.7 | 2.1 | 0.05 | 0.02 | 41.6 | 2.0 | | | 7.10 | 0.67 | 0.29 | 0.05 | 100.1 | 5 | *224* | *49* | *5877* |
| YN\* | S5157 | 21 | | | 43.5 | 1.8 | 0.09 | 0.00 | 47.0 | 1.4 | | | 5.26 | 0.31 | 0.44 | 0.04 | 96.7 | 3 | *32* | *6* | *6716* |

a) Silicate

\*reported by Fischer et al. (2015), trace elements in silicate re-analysed by ICPMS
\*\*other refers to trace elements present in the silicate but not of relevant to the present study. Group AV: Co and Ni; Group YN: Co, Cr, Mn, Ni, V; Group LRW: Au, Bi, Cr, Cu, Mn, Ni, Pb, Se, Sn, Te. The table was condensed for brevity and the complete data table including omitted trace elements is given in supplementary table S2.
\*\*\*time series, S-bearing starting material
italics indicate LA-ICP-MS measurements; all others are EPMA

| Group | Experiment | n EPMA | C | 1σ | Fe | 1σ | Mo | 1σ | Ni | 1σ | S | 1σ | W | 1σ | other* | Total |
|---|---|---|---|---|---|---|---|---|---|---|---|---|---|---|---|---|
| | | | | | Metal phase, wt. % | | | | | | | | | | | |
| AV | H3371a | 31 | | | 80.0 | 1.0 | 1.86 | 0.08 | 2.42 | 0.12 | 10.9 | 1.0 | 1.56 | 0.20 | 2.86 | 99.7 |
| AV | H3371b | | | | 82.7 | 3.3 | 3.28 | 1.04 | 2.32 | 0.36 | 0.25 | 0.09 | 3.81 | 0.76 | 2.73 | 94.7 |
| AV | H3372a | 30 | | | 83.2 | 0.6 | 3.05 | 0.06 | 2.02 | 0.07 | 6.28 | 0.10 | 2.06 | 0.22 | 2.59 | 99.2 |
| AV | H3372b | 30 | | | 95.1 | 0.5 | 0.70 | 0.16 | 1.22 | 0.07 | 0.04 | 0.02 | 1.09 | 0.26 | 2.35 | 100.5 |
| AV | H3586a | 30 | | | 91.3 | 0.5 | 1.09 | 0.08 | 1.56 | 0.06 | | | 1.61 | 0.19 | 2.69 | 98.2 |
| AV | H3586b | 30 | | | 93.4 | 0.6 | 0.83 | 0.15 | 1.26 | 0.05 | | | 1.22 | 0.21 | 2.39 | 99.2 |
| AV | Z1009b | 51 | | | 78.4 | 2.3 | 1.71 | 0.19 | 2.17 | 0.10 | 12.3 | 2.1 | 1.20 | 0.27 | 2.72 | 98.2 |
| AV | Z1011b | 50 | | | 90.7 | 1.5 | 1.10 | 0.08 | 1.74 | 0.06 | | | 1.87 | 0.19 | 3.46 | 99.0 |
| AV | Z822b | 30 | | | 93.9 | 1.4 | 0.85 | 0.10 | 1.31 | 0.06 | | | 1.61 | 0.20 | 2.82 | 100.5 |
| AV | Z859a | 29 | | | 95.0 | 1.0 | 0.88 | 0.09 | 1.33 | 0.05 | | | 1.63 | 0.25 | 3.28 | 102.1 |
| AV | Z865a | 31 | | | 94.4 | 0.9 | 0.87 | 0.13 | 1.39 | 0.05 | | | 1.55 | 0.24 | 3.01 | 101.3 |
| AV | Z916b | 31 | | | 95.1 | 0.6 | 0.84 | 0.14 | 1.32 | 0.06 | | | 1.48 | 0.21 | 2.29 | 101.0 |
| AV | Z929a | 30 | | | 93.4 | 0.7 | 0.84 | 0.13 | 1.33 | 0.05 | | | 1.50 | 0.24 | 2.49 | 99.6 |
| AV | Z950a | 31 | | | 93.5 | 2.1 | 0.80 | 0.12 | 1.31 | 0.07 | | | 1.48 | 0.24 | 3.10 | 100.4 |
| AV | Z970a | 40 | | | 94.1 | 0.5 | 0.84 | 0.08 | 1.35 | 0.05 | | | 1.51 | 0.16 | 2.94 | 100.7 |
| AV | Z970b | 40 | | | 93.8 | 0.9 | 0.85 | 0.18 | 1.37 | 0.06 | | | 1.52 | 0.24 | 3.00 | 100.5 |
| ESJ | A962b | 32 | 5.00 | 0.29 | 89.9 | 0.9 | 1.85 | 0.07 | | | | | 2.03 | 0.16 | | 98.8 |
| ESJ | A961a-L | 47 | 3.33 | 0.17 | 93.7 | 0.6 | 1.55 | 0.07 | | | | | 1.85 | 0.09 | | 100.5 |
| ESJ | A961a-R | 48 | 1.32 | 0.07 | 95.3 | 0.5 | 1.91 | 0.15 | | | | | 1.90 | 0.13 | | 100.4 |
| ESJ | A961b-L | 17 | 1.94 | 0.14 | 92.1 | 0.8 | 1.55 | 0.12 | | | | | 1.68 | 0.14 | | 97.2 |
| ESJ | A961b-R | 38 | 0.08 | 0.07 | 95.4 | 1.0 | 1.97 | 0.04 | | | | | 1.99 | 0.06 | | 99.5 |
| LRW | A81 | 56 | | | 83.3 | 1.8 | | | 4.45 | 0.13 | | | 1.67 | 0.13 | 9.8 | 99.3 |
| LRW | A73 | 13 | | | 82.6 | 2.9 | | | 4.70 | 0.13 | 5.54 | 2.03 | 0.95 | 0.20 | 5.1 | 98.9 |
| LRW | A74 | 51 | | | 77.0 | 2.6 | | | 4.45 | 0.22 | 8.60 | 1.63 | 0.68 | 0.15 | 6.7 | 97.4 |
| LRW | H2603 | 12 | | | 62.9 | 0.9 | | | 4.16 | 0.14 | 7.97 | 0.51 | 13.91 | 0.43 | 13.7 | 102.6 |
| LRW | V378 | 10 | | | 80.6 | 0.5 | | | 5.33 | 0.10 | 10.9 | 0.4 | 0.12 | 0.09 | 1.9 | 98.9 |
| LRW | H2617 | 24 | | | 80.4 | 0.7 | | | 5.25 | 0.18 | 10.9 | 0.5 | 0.06 | 0.06 | 1.8 | 98.3 |
| LRW | H2775 | 20 | | | 76.0 | 1.4 | | | 5.08 | 0.21 | 10.4 | 1.5 | 0.73 | 0.21 | 5.7 | 97.9 |
| LRW | H2735 | 20 | | | 69.2 | 0.7 | | | 7.25 | 0.35 | 11.3 | 0.8 | 0.49 | 0.13 | 6.8 | 95.1 |
| LRW | H2748 | 20 | | | 75.6 | 1.0 | | | 5.80 | 0.16 | 11.0 | 0.6 | 0.62 | 0.12 | 6.1 | 99.1 |
| LRW | H2749 | 15 | | | 74.4 | 1.6 | | | 6.02 | 0.16 | 11.2 | 1.1 | 0.77 | 0.18 | 6.9 | 99.2 |
| LRW | H2755 | 21 | | | 74.8 | 0.8 | | | 6.64 | 0.13 | 10.2 | 0.4 | 0.66 | 0.09 | 6.2 | 98.5 |

| | | | | | | | | | | | | |
|---|---|---|---|---|---|---|---|---|---|---|---|---|
| LRW | A129 | 23 | 70.1 | 0.9 | 4.09 | 0.17 | 14.6 | 0.5 | 0.90 | 0.09 | 9.7 | 99.4 |
| LRW | 4A*** | 63 | 62.1 | 2.6 | 3.63 | 0.53 | 21.3 | 2.1 | 0.27 | 0.26 | 11.3 | 98.5 |
| LRW | 4At2*** | 10 | 61.2 | 1.6 | 3.66 | 0.48 | 21.1 | 1.4 | 0.20 | 0.10 | 12.4 | 98.6 |
| LRW | 4At3*** | 18 | 58.1 | 4.8 | 3.65 | 0.96 | 21.0 | 3.8 | 0.09 | 0.10 | 15.0 | 97.7 |
| LRW | 4At4*** | 26 | 60.5 | 3.8 | 4.14 | 1.35 | 19.6 | 4.2 | 0.15 | 0.35 | 13.7 | 98.1 |
| LRW | 4At5*** | 30 | 62.5 | 2.2 | 3.77 | 0.52 | 20.1 | 1.8 | 0.33 | 0.16 | 12.2 | 98.9 |
| YN | S5700 | 6 | 89.3 | 0.7 | 6.44 | 0.22 | | | 1.87 | 0.11 | 3.9 | 101.5 |
| YN | S5703 | 13 | 92.3 | 1.7 | 4.30 | 0.79 | | | 1.46 | 0.25 | 3.9 | 102.0 |
| YN | S5706 | 31 | 90.7 | 0.8 | 5.91 | 0.10 | | | 1.91 | 0.12 | 2.9 | 101.4 |
| YN | H3346 | 13 | 82.2 | 0.8 | 6.27 | 0.11 | 3.79 | 0.18 | 2.05 | 0.06 | 5.0 | 99.2 |
| YN* | S4818 | 52 | 86.4 | 2.8 | 6.24 | 2.27 | | | 1.59 | 0.57 | 6.3 | 100.6 |
| YN* | S4842 | 10 | 85.3 | 1.2 | 7.09 | 0.89 | | | 2.31 | 0.39 | 5.6 | 100.4 |
| YN* | S4858 Fo95-A | 38 | 84.7 | 0.7 | 4.76 | 0.43 | | | 1.32 | 0.20 | 10.3 | 101.1 |
| YN* | S4858 Fo70-B | 39 | 86.7 | 1.7 | 7.19 | 0.59 | | | 2.30 | 0.24 | 4.2 | 100.4 |
| YN* | S4933 | 18 | 91.6 | 0.7 | 3.82 | 0.24 | | | 1.24 | 0.16 | 4.1 | 100.8 |
| YN | S5056-Spoor | 20 | 89.2 | 1.3 | 4.95 | 0.06 | 2.16 | 0.56 | 1.62 | 0.09 | 3.5 | 101.4 |
| YN* | S5069 | 26 | 85.6 | 1.5 | 6.55 | 0.89 | | | 2.29 | 0.31 | 5.1 | 99.5 |
| YN* | S5134 | 9 | 86.4 | 0.7 | 7.17 | 0.10 | | | 2.17 | 0.11 | 3.9 | 99.6 |
| YN* | S5139 | 98 | 84.8 | 0.6 | 3.89 | 0.04 | | | 1.09 | 0.07 | 10.6 | 100.4 |
| YN* | S5140 | 27 | 84.3 | 0.4 | 7.47 | 0.06 | | | 2.58 | 0.08 | 4.7 | 99.0 |
| YN* | S5157 | 14 | 86.3 | 1.9 | 5.02 | 1.35 | | | 1.51 | 0.51 | 4.6 | 97.4 |

b) Metal

Table 3: Values of $\varepsilon_i^j$ and $\gamma^0$ used and determined in this study, at 1873 K. Values in italics: no published value; negligible interaction assumed. References: TS, Values redetermined in this study (section 5.7); T11, Tuff et al. (2011); W12*, from the online erratum to Wade et al. (2012) at http://norris.org.au/expet/metalact/?p=about; W14, Wood et al. (2014); all others from J.S.P.S (1988).

| $\varepsilon_i^j$ | C | Mo | Ni | O | S | Si | W |
|---|---|---|---|---|---|---|---|
| C | 12.83 | -7.03 (TS) | 2.31 | -20.41 | 6.09 | 9.77 | -7.38 (TS) |
| Mo | | 3.94 | 0 | 1.26 | 2.27 (W14) | 22 (T11) | - |
| Ni | | | 0.119 | 1.37 | 2.17 (W14) | 7.5 (T11) | 0 |
| O | | | | -10.51 | -17.17 | -7.15 | 4.16 |
| S | | | | | -5.66 | 8.96 | 6.47 (W14) |
| Si | | | | | | 8.6 (T11) | 18.3 (T11) |
| W | | | | | | | 0 |
| $\gamma^0$ | | 1 | | | | | 3.0 (W12*) |

Table 4: Calculated log $fO_2$ ΔIW (with and without activity correction for the metal; see text), mole fractions, activity coefficients ($\gamma_M$), log $D_M$, log $K_D^M$, and log $K^M$, from the experiments listed in Table 2a, b.

| Experiment | log $fO_2$ ΔIW ideal | log $fO_2$ ΔIW activity-corr. | $X_{FeO}$ | $X_{MoO2}$ 1E-05 | $X_{WO3}$ 1E-05 | $X_{Fe}$ | $X_{Mo}$ 1E-03 | $X_W$ 1E-03 | $\gamma_{Fe}$ | $\gamma_{Mo}$ | $\gamma_W$ | log $D_{Mo}$ | log $D_W$ | log $K_D^{Mo}$ | log $K_D^W$ | log $K^{Mo}$ | log $K^W$ |
|---|---|---|---|---|---|---|---|---|---|---|---|---|---|---|---|---|---|
| H3371a | -2.59 | -2.66 | 0.04 | 1.79 | 4.76 | 0.75 | 10.20 | 4.44 | 1.10 | 1.56 | 5.82 | 2.76 | 1.97 | 0.17 | -1.91 | 0.28 | -1.26 |
| H3371b | -2.91 | -2.91 | 0.03 | 0.45 | 1.28 | 0.90 | 20.87 | 12.64 | 1.00 | 1.08 | 2.20 | 3.67 | 2.99 | 0.75 | -1.38 | 0.79 | -1.03 |
| H3372a | -2.46 | -2.50 | 0.05 | 1.41 | 4.66 | 0.82 | 17.39 | 6.13 | 1.05 | 1.38 | 4.52 | 3.09 | 2.12 | 0.63 | -1.57 | 0.73 | -0.98 |
| H3372b | -2.72 | -2.72 | 0.04 | 0.92 | 1.32 | 0.95 | 4.05 | 3.29 | 1.00 | 1.08 | 2.53 | 2.64 | 2.40 | -0.07 | -1.68 | -0.04 | -1.28 |
| H3586a | -2.82 | -2.82 | 0.04 | 0.21 | 1.61 | 0.94 | 6.53 | 5.04 | 1.00 | 1.03 | 2.27 | 3.50 | 2.50 | 0.67 | -1.74 | 0.69 | -1.38 |
| H3586b | -2.97 | -2.97 | 0.03 | 0.17 | 0.91 | 0.95 | 4.89 | 3.75 | 1.00 | 1.04 | 2.32 | 3.45 | 2.61 | 0.49 | -1.84 | 0.50 | -1.47 |
| Z1009b | -2.51 | -2.62 | 0.04 | 1.04 | 7.29 | 0.74 | 9.33 | 3.42 | 1.13 | 1.72 | 7.55 | 2.95 | 1.67 | 0.44 | -2.10 | 0.57 | -1.38 |
| Z1011b | -2.80 | -2.80 | 0.04 | 0.15 | 0.47 | 0.92 | 6.47 | 5.77 | 1.00 | 1.21 | 2.65 | 3.64 | 3.09 | 0.83 | -1.12 | 0.92 | -0.70 |
| Z822b | -2.76 | -2.76 | 0.04 | 0.17 | 0.71 | 0.94 | 4.94 | 4.89 | 1.00 | 1.16 | 2.58 | 3.46 | 2.84 | 0.70 | -1.31 | 0.76 | -0.90 |
| Z859a | -2.56 | -2.56 | 0.05 | 0.19 | 0.74 | 0.93 | 4.97 | 4.82 | 1.00 | 1.27 | 2.69 | 3.42 | 2.81 | 0.86 | -1.02 | 0.97 | -0.59 |
| Z865a | -2.65 | -2.65 | 0.04 | 0.24 | 0.84 | 0.93 | 4.98 | 4.66 | 1.00 | 1.20 | 2.46 | 3.31 | 2.75 | 0.66 | -1.22 | 0.74 | -0.83 |
| Z916b | -2.81 | -2.81 | 0.04 | 0.35 | 1.65 | 0.95 | 4.88 | 4.51 | 1.00 | 1.09 | 2.38 | 3.15 | 2.44 | 0.33 | -1.78 | 0.37 | -1.41 |
| Z929a | -2.60 | -2.60 | 0.05 | 0.27 | 1.19 | 0.95 | 4.97 | 4.61 | 1.00 | 1.13 | 2.31 | 3.27 | 2.59 | 0.67 | -1.32 | 0.72 | -0.95 |
| Z950a | -2.49 | -2.49 | 0.05 | 0.55 | 1.20 | 0.93 | 4.64 | 4.49 | 1.00 | 1.30 | 2.55 | 2.92 | 2.57 | 0.43 | -1.16 | 0.55 | -0.76 |
| Z970a | -2.58 | -2.58 | 0.05 | 0.38 | 0.80 | 0.94 | 4.88 | 4.57 | 1.00 | 1.26 | 2.45 | 3.11 | 2.76 | 0.54 | -1.11 | 0.64 | -0.72 |
| Z970b | -2.63 | -2.63 | 0.05 | 0.29 | 0.61 | 0.94 | 4.93 | 4.59 | 1.00 | 1.24 | 2.45 | 3.23 | 2.88 | 0.59 | -1.07 | 0.69 | -0.68 |
| A962b | -2.13 | -1.87 | 0.07 | 0.02 | 0.35 | 0.78 | 9.37 | 5.38 | 0.75 | 0.16 | 0.42 | 4.64 | 3.18 | 2.51 | -0.01 | 1.97 | -0.01 |
| A961a-L | -2.67 | -2.56 | 0.04 | 0.04 | 1.55 | 0.85 | 8.14 | 5.09 | 0.88 | 0.32 | 0.85 | 4.32 | 2.52 | 1.65 | -1.50 | 1.26 | -1.39 |
| A961a-R | -2.60 | -2.58 | 0.05 | 0.11 | 3.97 | 0.92 | 10.80 | 5.59 | 0.98 | 0.67 | 1.83 | 4.00 | 2.15 | 1.39 | -1.75 | 1.24 | -1.47 |
| A961b-L | -2.67 | -2.63 | 0.04 | 0.06 | 2.07 | 0.90 | 8.81 | 4.98 | 0.95 | 0.52 | 1.42 | 4.18 | 2.38 | 1.51 | -1.63 | 1.27 | -1.42 |
| A961b-R | -2.66 | -2.66 | 0.05 | 0.14 | 4.71 | 0.98 | 11.77 | 6.20 | 1.00 | 1.02 | 2.84 | 3.91 | 2.12 | 1.26 | -1.87 | 1.26 | -1.41 |
| A81 | -2.23 | -2.23 | 0.07 | | 4.11 | 0.90 | | 5.47 | 1.00 | | 2.52 | | 2.12 | | -1.22 | | -0.81 |
| A73 | -2.32 | -2.33 | 0.06 | | 24.0 | 0.83 | | 2.90 | 1.01 | | 4.54 | | 1.08 | | -2.39 | | -1.75 |
| A74 | -1.99 | -2.03 | 0.08 | | 38.7 | 0.77 | | 2.07 | 1.05 | | 6.65 | | 0.73 | | -2.25 | | -1.49 |
| H2603 | -2.67 | -2.67 | 0.03 | | 47.2 | 0.69 | | 46.09 | 1.00 | | 5.01 | | 1.99 | | -2.02 | | -1.32 |
| V378 | -2.70 | -2.75 | 0.03 | | 1.42 | 0.76 | | 0.34 | 1.06 | | 5.61 | | 1.38 | | -2.67 | | -1.99 |
| H2617 | -2.71 | -2.76 | 0.03 | | 0.65 | 0.76 | | 0.17 | 1.06 | | 6.06 | | 1.43 | | -2.64 | | -1.94 |
| H2775 | -2.38 | -2.49 | 0.05 | | 0.60 | 0.73 | | 2.14 | 1.13 | | 8.61 | | 2.55 | | -1.01 | | -0.24 |
| H2735 | -2.18 | -2.33 | 0.06 | | 8.51 | 0.68 | | 1.46 | 1.19 | | 7.70 | | 1.23 | | -2.03 | | -1.37 |

| | | | | | | | | | | | |
|---|---|---|---|---|---|---|---|---|---|---|---|
| H2748 | -2.77 | -2.89 | 0.03 | 1.22 | 0.71 | 1.78 | 1.15 | 6.79 | 2.16 | -1.98 | -1.34 |
| H2749 | -2.56 | -2.69 | 0.04 | 2.90 | 0.70 | 2.22 | 1.15 | 6.94 | 1.88 | -1.96 | -1.31 |
| H2755 | -2.44 | -2.54 | 0.04 | 2.40 | 0.72 | 1.92 | 1.13 | 6.65 | 1.90 | -1.76 | -1.09 |
| A129 | -1.78 | -2.07 | 0.08 | 34.4 | 0.64 | 2.51 | 1.40 | 18.90 | 0.86 | -1.81 | -0.97 |
| 4A* | -1.72 | -2.19 | 0.08 | 96.5 | 0.56 | 0.74 | 1.73 | 50.83 | -0.12 | -2.70 | -1.70 |
| 4At2* | -1.68 | -2.26 | 0.08 | 178.0 | 0.55 | 0.54 | 1.95 | 58.18 | -0.52 | -3.04 | -2.14 |
| 4At3* | -1.60 | -2.39 | 0.08 | 179.1 | 0.52 | 0.24 | 2.47 | 81.47 | -0.87 | -3.27 | -2.54 |
| 4At4* | -1.66 | -2.27 | 0.08 | 251.5 | 0.54 | 0.41 | 2.03 | 54.15 | -0.79 | -3.27 | -2.46 |
| 4At5* | -2.26 | -2.76 | 0.04 | 102.5 | 0.56 | 0.90 | 1.77 | 46.49 | -0.06 | -3.44 | -2.52 |
| S5700 | -2.35 | -2.35 | 0.06 | 7.18 | 0.88 | 5.60 | 0.99 | 2.63 | 1.89 | -1.64 | -1.21 |
| S5703 | -1.88 | -1.88 | 0.10 | 10.9 | 0.89 | 4.27 | 1.00 | 2.81 | 1.59 | -1.23 | -0.78 |
| S5706 | -1.86 | -1.86 | 0.11 | 11.7 | 0.89 | 5.73 | 1.00 | 2.27 | 1.69 | -1.10 | -0.74 |
| H3346 | -2.58 | -2.57 | 0.04 | 2.18 | 0.80 | 6.08 | 0.99 | 3.90 | 2.44 | -1.43 | -0.82 |
| S4818 | -2.76 | -2.73 | 0.04 | 3.32 | 0.84 | 4.70 | 0.97 | 4.37 | 2.15 | -1.99 | -1.30 |
| S4842 | -2.54 | -2.52 | 0.05 | 1.13 | 0.85 | 6.96 | 0.97 | 3.39 | 2.79 | -1.02 | -0.45 |
| S4858Fo95-A | -3.13 | -3.05 | 0.02 | 0.13 | 0.79 | 3.74 | 0.92 | 9.54 | 3.45 | -1.24 | -0.16 |
| S4858Fo70-B | -1.88 | -1.88 | 0.10 | 6.86 | 0.85 | 6.89 | 1.00 | 2.44 | 2.00 | -0.81 | -0.43 |
| S4933 | -2.57 | -2.56 | 0.05 | 0.55 | 0.89 | 3.66 | 0.99 | 3.52 | 2.82 | -1.04 | -0.47 |
| S5056-Spoor | -2.00 | -2.01 | 0.09 | 6.59 | 0.86 | 4.75 | 1.00 | 3.00 | 1.86 | -1.15 | -0.67 |
| S5069 | -2.80 | -2.78 | 0.03 | 0.53 | 0.86 | 7.01 | 0.99 | 3.00 | 3.12 | -1.07 | -0.58 |
| S5134 | -1.99 | -1.99 | 0.09 | 9.09 | 0.87 | 6.59 | 1.00 | 2.35 | 1.86 | -1.13 | -0.76 |
| S5139 | -3.65 | -3.58 | 0.01 | 0.07 | 0.79 | 3.09 | 0.92 | 11.32 | 3.63 | -1.85 | -0.69 |
| S5140 | -2.51 | -2.50 | 0.05 | 5.90 | 0.86 | 7.99 | 0.99 | 2.69 | 2.13 | -1.63 | -1.18 |
| S5157 | -2.74 | -2.73 | 0.04 | 0.89 | 0.88 | 4.68 | 0.99 | 3.08 | 2.72 | -1.39 | -0.89 |

*time series, S-bearing starting material

Table 5: Restrictions in major element compositions (wt. %) applied to the silicate melt compositions used when modelling partitioning data.

| Oxide | Limit for $K^W$ | Limit for $K^{Mo}$ |
|---|---|---|
| MgO | > 12 | |
| SiO$_2$ | < 55 | |
| Al$_2$O$_3$ | < 20 | |
| CaO | < 10 | |
| Na$_2$O | < 2 | < 2 |

## 9. Figures

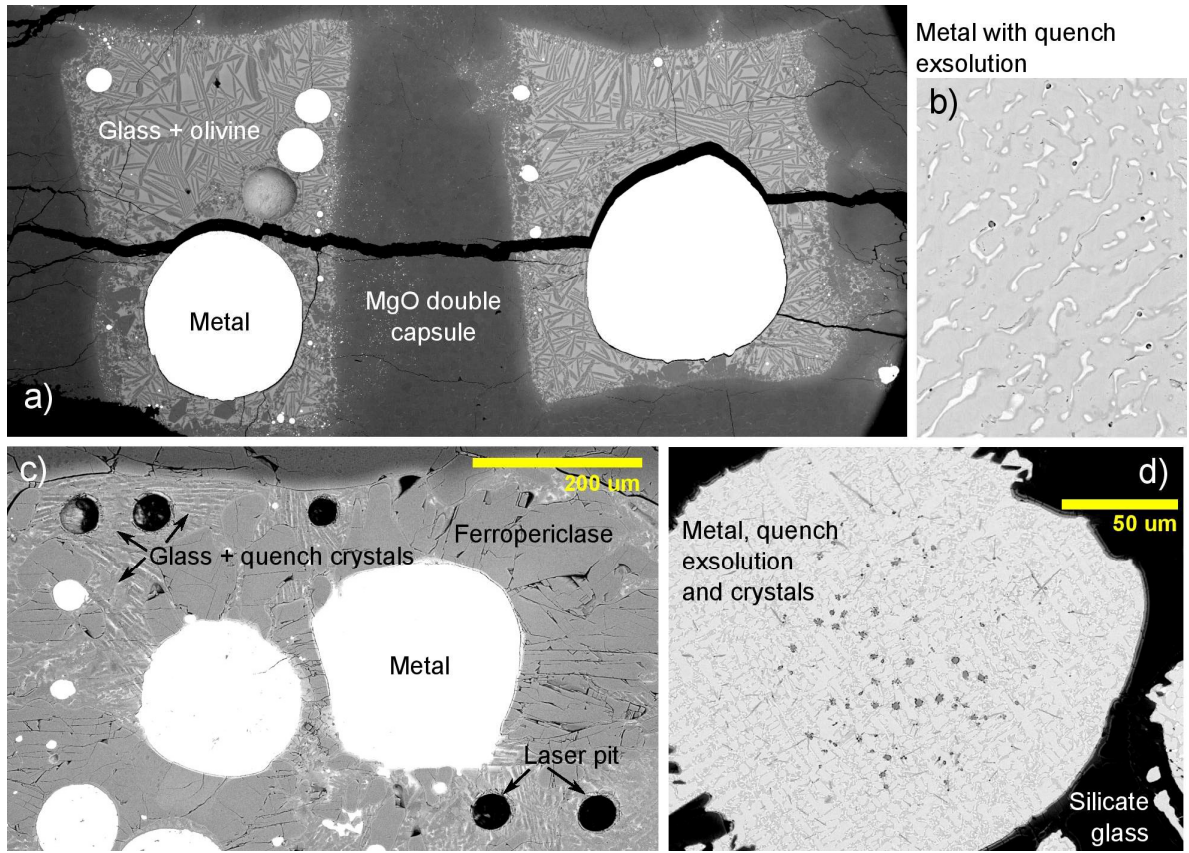

Fig. 1: Back scattered electron images of some representative samples. a) Double MgO capsule experiment performed in a piston cylinder at 1.5 GPa: A961b-L and A961b-R, which contain added graphite. Spinifex olivine quench crystals are seen. b) High contrast and high magnification image of the metal phase of A961b-L, showing exsolution of two Fe-rich metal phases. c) experiment S4858B, performed in a multi-anvil press at 25 GPa in a double MgO capsule (other experiment not shown). Large ferropericlase crystals, precipitated during the experiment, are clearly visible, along with fine quench crystals and ablation pits from LA-ICP-MS analysis. d) High contrast image of the metal phase of multi-anvil experiment S5056 (25 GPa), showing quench features: immiscibility of two metals and the exsolution of silica crystals.

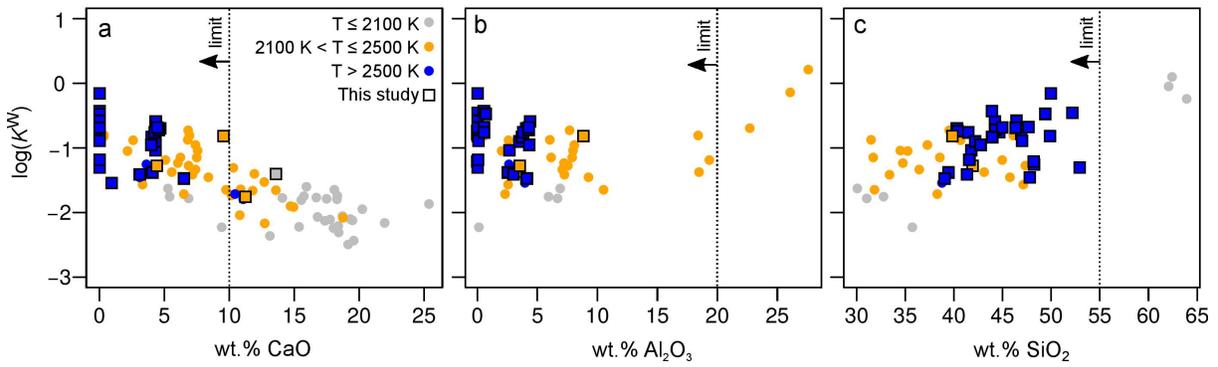

*Fig. 2: log($K^W$) as a function of a) CaO, with compositions restricted in all components except CaO; b) $Al_2O_3$, with composition restricted in all components except $Al_2O_3$; c) $SiO_2$, where composition is restricted in all components except $SiO_2$ and MgO. Compositional restrictions are given in Table 4, where the limits are shown here by dotted lines. Different colours represent temperature ranges of the experiments, where circles are published data and squares with black outlines show new data from this study.*

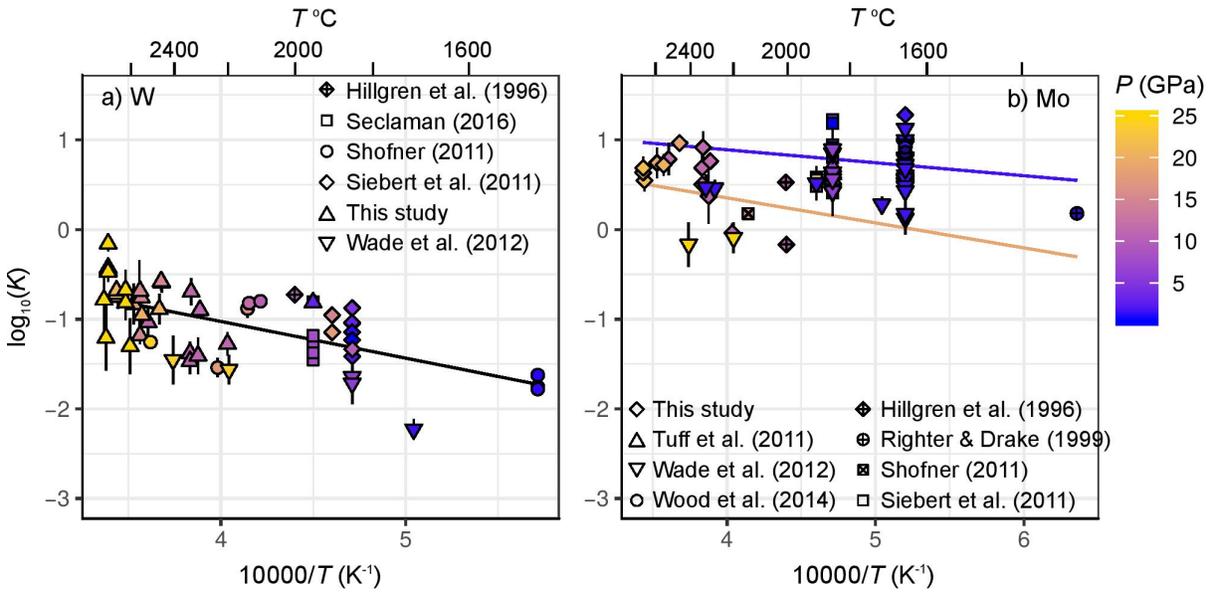

*Fig. 3: a) log($K^W$) and b) log($K^{Mo}$) as a function of inverse temperature (10000/T). Colour scale indicates experimental pressure, line shows fitted model prediction, where the two lines in (b) represent the model fit for 0 and 20 GPa. Symbols indicate individual data sources. Where no error bar is shown, it is either smaller than the symbol or is not provided (in the case of some published values).*

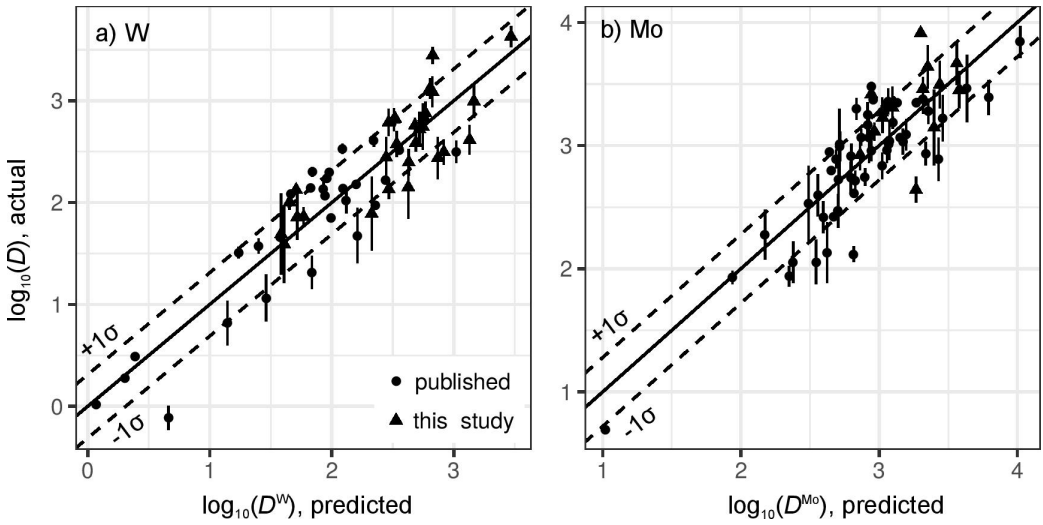

*Fig. 4: Measured molar log(D) vs. predicted molar log(D), for a) W and b) Mo. Predicted values are calculated using eq. 8. Solid line shows unity, with dashed lines representing a 1 RMS departure from unity.*

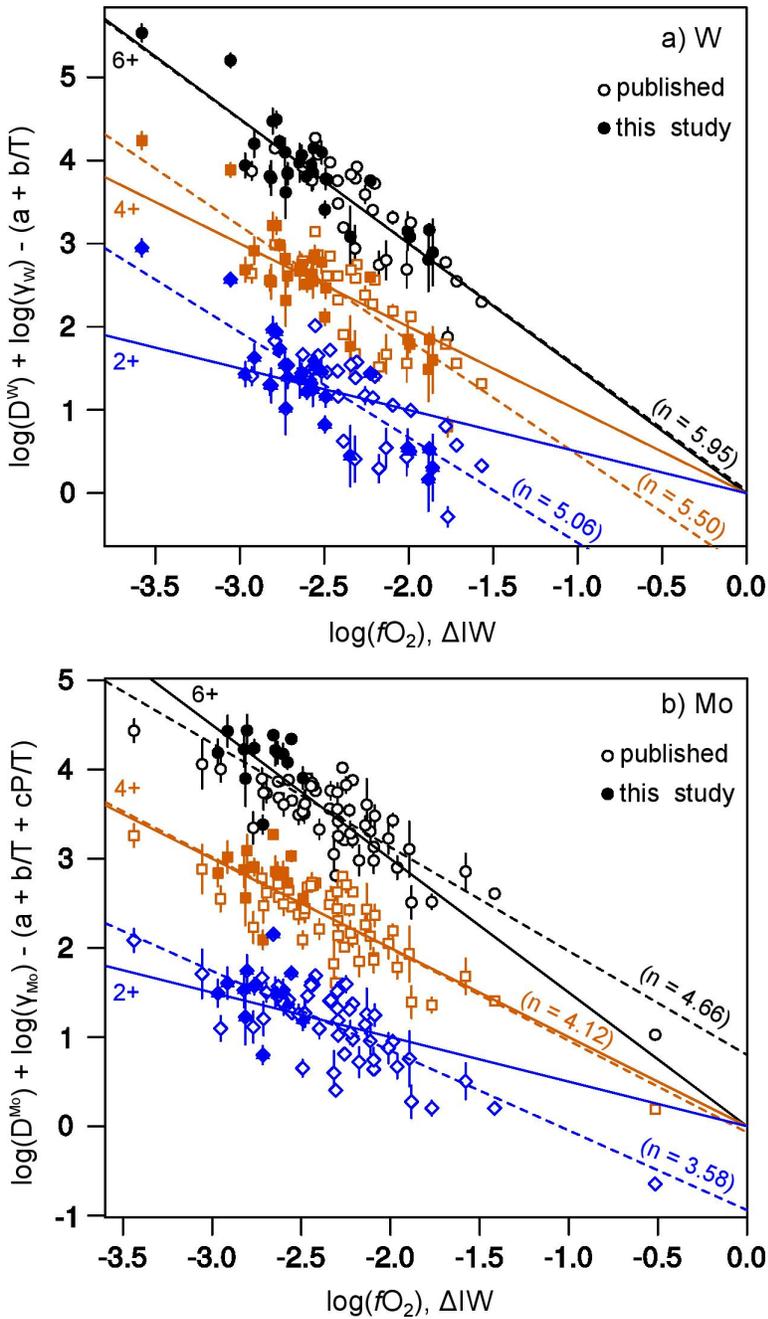

*Fig. 5: log(D) adjusted for trace metal activity, the known intercept and the effect of P and T, as a function of log(fO$_2$), ΔIW (eq. 8; see text), for a) W and b) Mo. Data expressed in this way show the link between valence n, D, and fO$_2$, where the slope is equal to $-\frac{n}{4}$. Solid lines show the predicted gradient (i.e. $-\frac{n}{4}$); dashed lines show the least squares gradient fitted to the data. Black (highest values) shows the trend for n=6, orange for n=4 and blue (lowest values) for n=2. Open symbols are published data filtered as described in the text; closed symbols are new data. The fitted slopes correspond to n values that are indicated on the plot, and the data correspond to the following mismatches to the predicted values (solid line): a) W: n=6, RMS=0.31; n=4, RMS=0.35; n=2, RMS=0.45; b) Mo: n=6, RMS=0.32; n=4, RMS=0.28; n=2, RMS=0.34.*

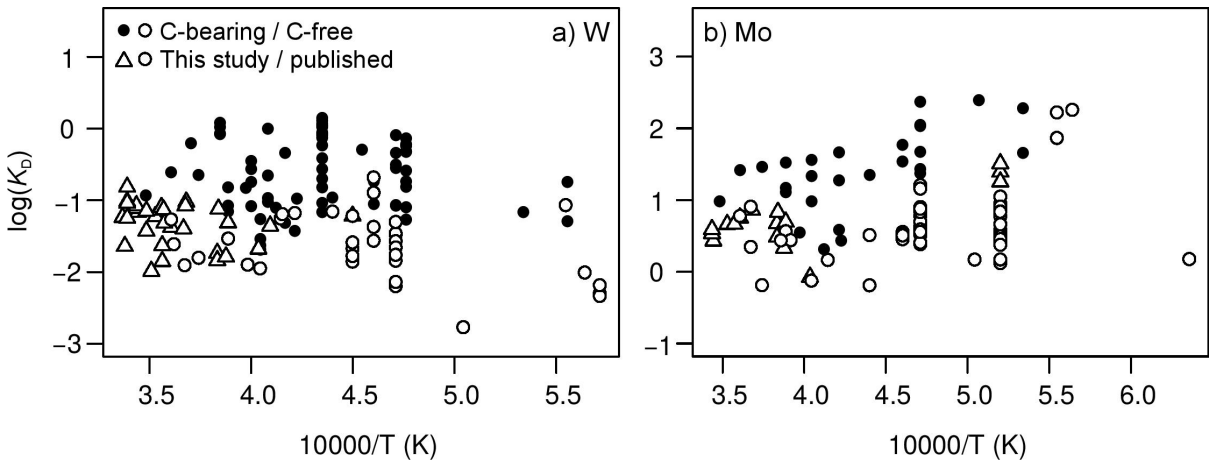

*Fig. 6: $K_D$ of C-bearing (graphite capsule; filled symbols) and C-free (non-graphite capsule; open symbols) experiments from this study (triangles) and published studies (circles) for a) W and b) Mo as a function of inverse temperature.*

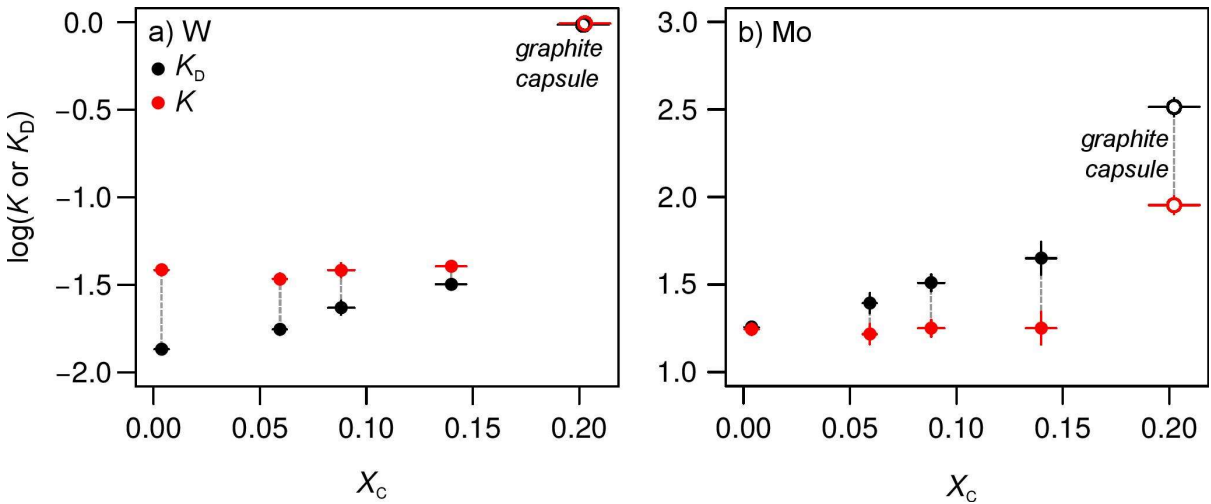

*Fig. 7: K (red) and $K_D$ (black) for a) W and b) Mo as a function of mole fraction of C in the metal $X_C$ for five experiments performed at 1.5 GPa and 1923 K to determine the effect of carbon content of the metal on partitioning. Filled symbols show data points from experiments in MgO capsules; open symbol in a graphite capsule. The activity correction made in the calculation of K utilised the carbon term $\varepsilon_C^M$.*

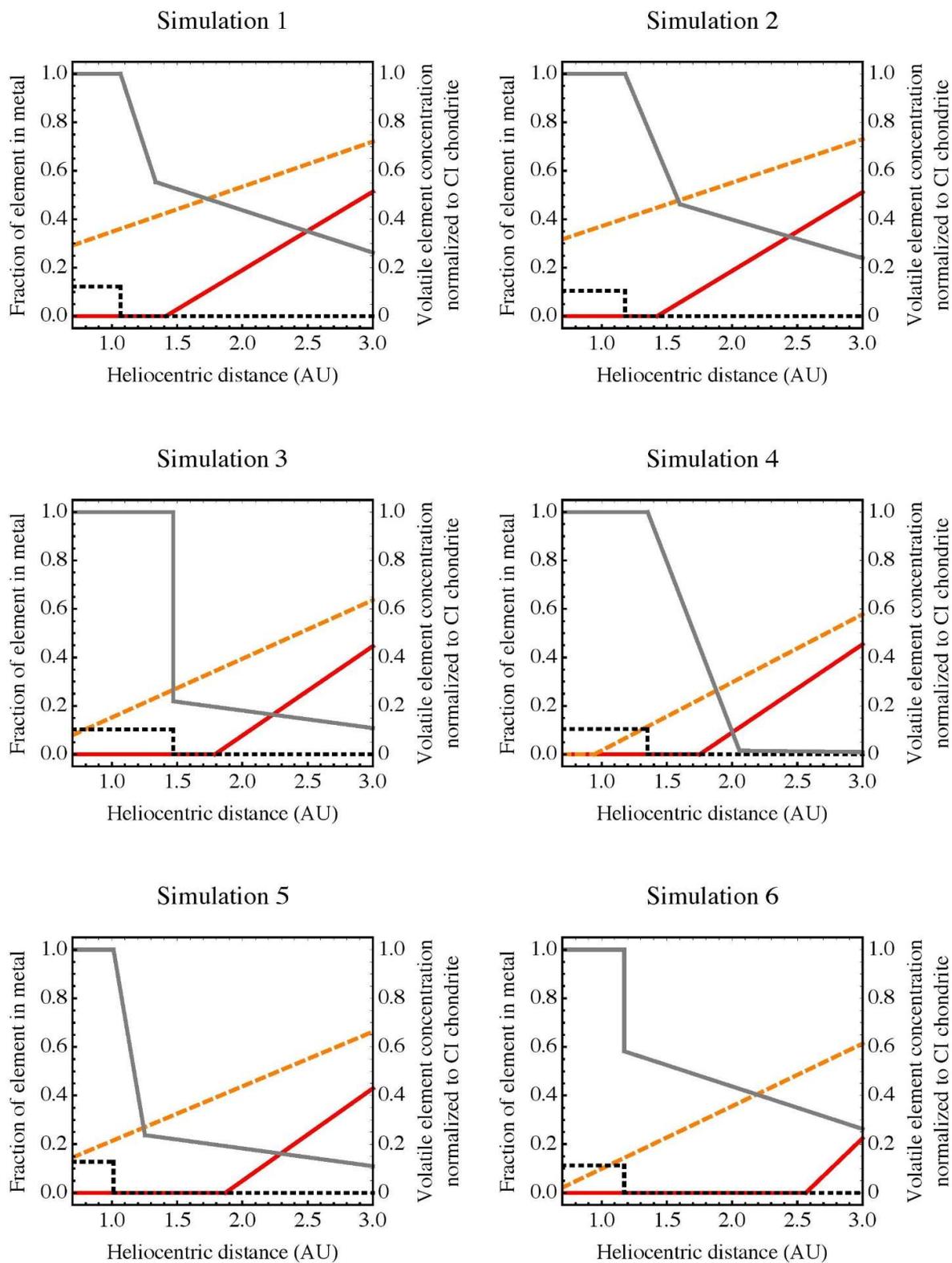

*Fig. 8: Oxidation and volatile element concentration gradients for each of the six accretion simulations as a function of heliocentric distance that result in the final tracked element abundances for the Earth-like planet closely matching Earth, assuming it has a high C mantle abundance (786 ± 308 ppm; Marty, 2012; see supplementary Fig. S6 for how the gradients change when considering*

*a low C mantle abundance). The fraction of Fe (grey, solid) and Si (black, dotted) in metal within each original planetesimal and embryo in the N-body simulations defines the oxidation gradient and is shown as a function of radial distance from the Sun using the left-hand ordinate. The concentration of S (orange, dashed) and C (red, solid) within each initial planetesimal and embryo relative to the concentration in CI chondrites (S: 5.35 wt.% and C: 3.45 wt.%) is shown using the right-hand ordinate. Planetesimals delivered from the giant planet forming region (>4.5 AU) are fully oxidized and contain CI chondrite-like concentrations of S and C.*

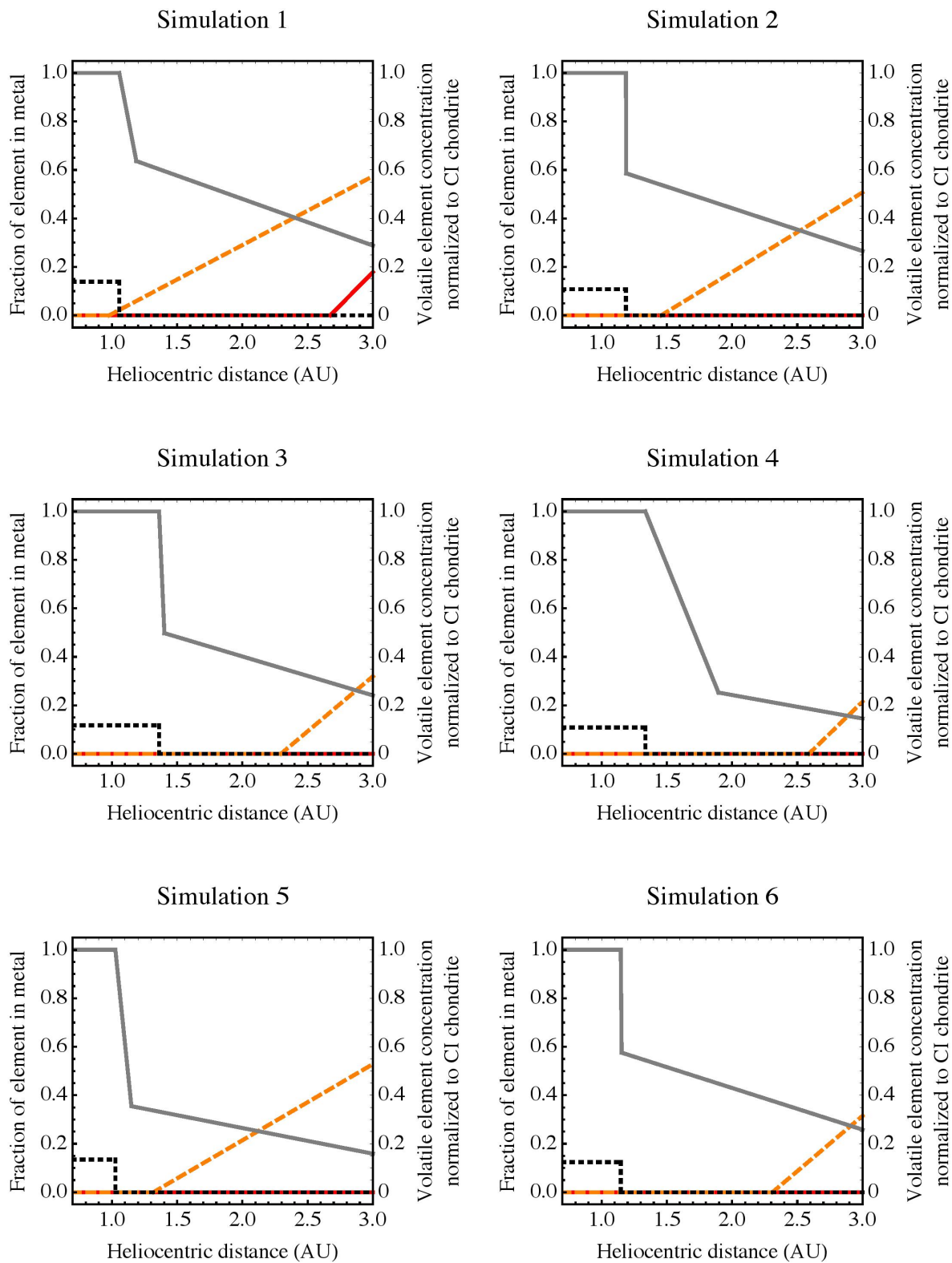

*Fig. 9: Similar to Fig. 8, oxidation and volatile element concentration gradients for each of the six accretion simulations as a function of heliocentric distance that result in the final tracked element abundances for the Earth-like planet closely matching Earth, assuming it has a low C mantle abundance (66 ± 21 ppm; Halliday, 2013). The fraction of Fe (grey, solid) and Si (black, dotted) in*

*metal within each original planetesimal and embryo in the N-body simulation defines the oxidation gradient and is shown as a function of radial distance from the Sun using the left-hand ordinate. The concentration of S (orange, dashed) and C (red, solid) within each initial planetesimal and embryo relative to the concentration in CI chondrites (S: 5.35 wt.% and C: 3.45 wt.%) is shown using the right-hand ordinate. Planetesimals delivered from the giant planet forming region are fully oxidized and contain CI chondrite-like concentrations of S and C.*

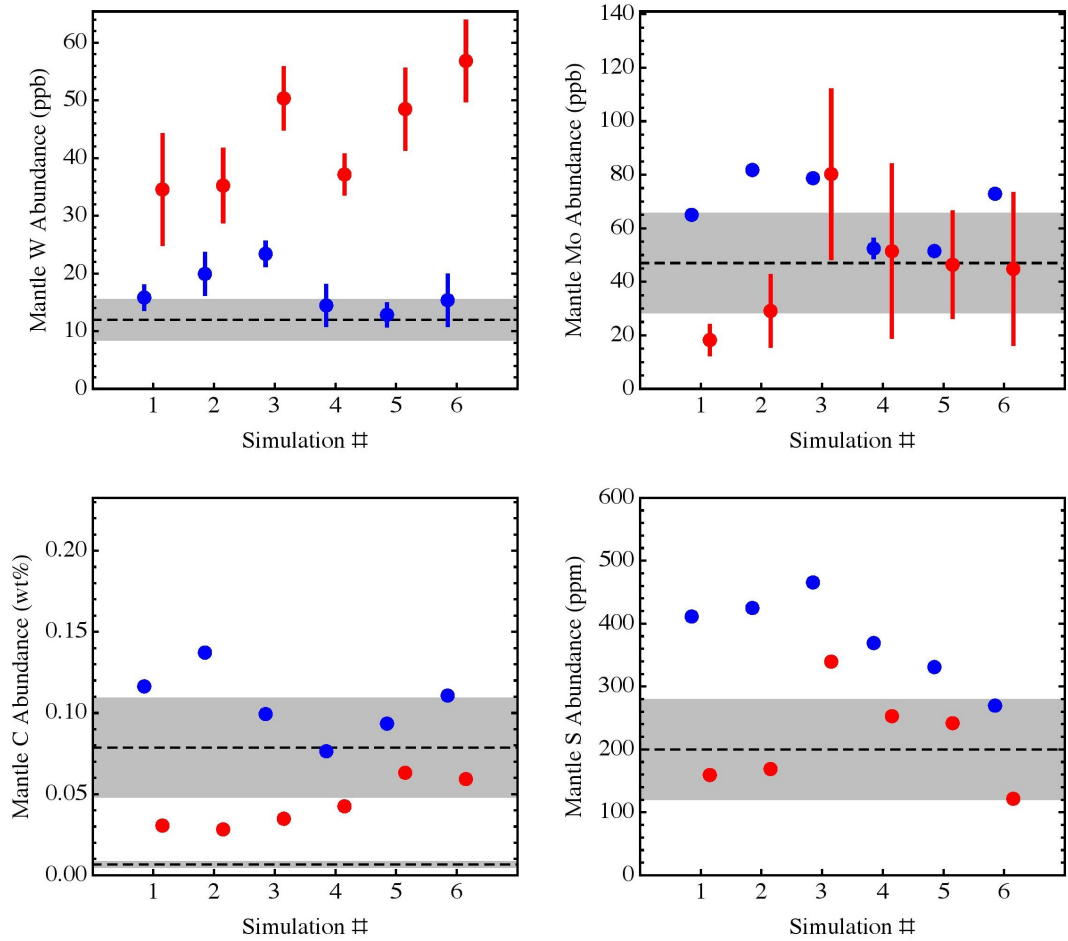

*Fig. 10: The best fit mantle abundances of W, Mo, C and S for the simulated Earth-like planets from accretion simulations 1-6 when considering both a high (blue) and a low (red) C concentration estimate for Earth's mantle. Error bars of the best fit mantle uncertainties, when large enough to be visible, are the one-sigma laboratory measurement partition coefficient uncertainties propagated through the numerical model into absolute mantle abundances. For W, Mo, and S, the horizontal black dashed lined is the estimated mantle abundance with a corresponding shaded horizontal band indicating the one-sigma uncertainties of that estimate (Palme & O'Neill, 2014). For C, the upper dashed line and shaded band correspond to the mantle abundance and uncertainty estimate of Marty (2012) whereas the lower dashed line and shaded band correspond to the mantle C abundance and uncertainty estimate of Halliday (2013).*

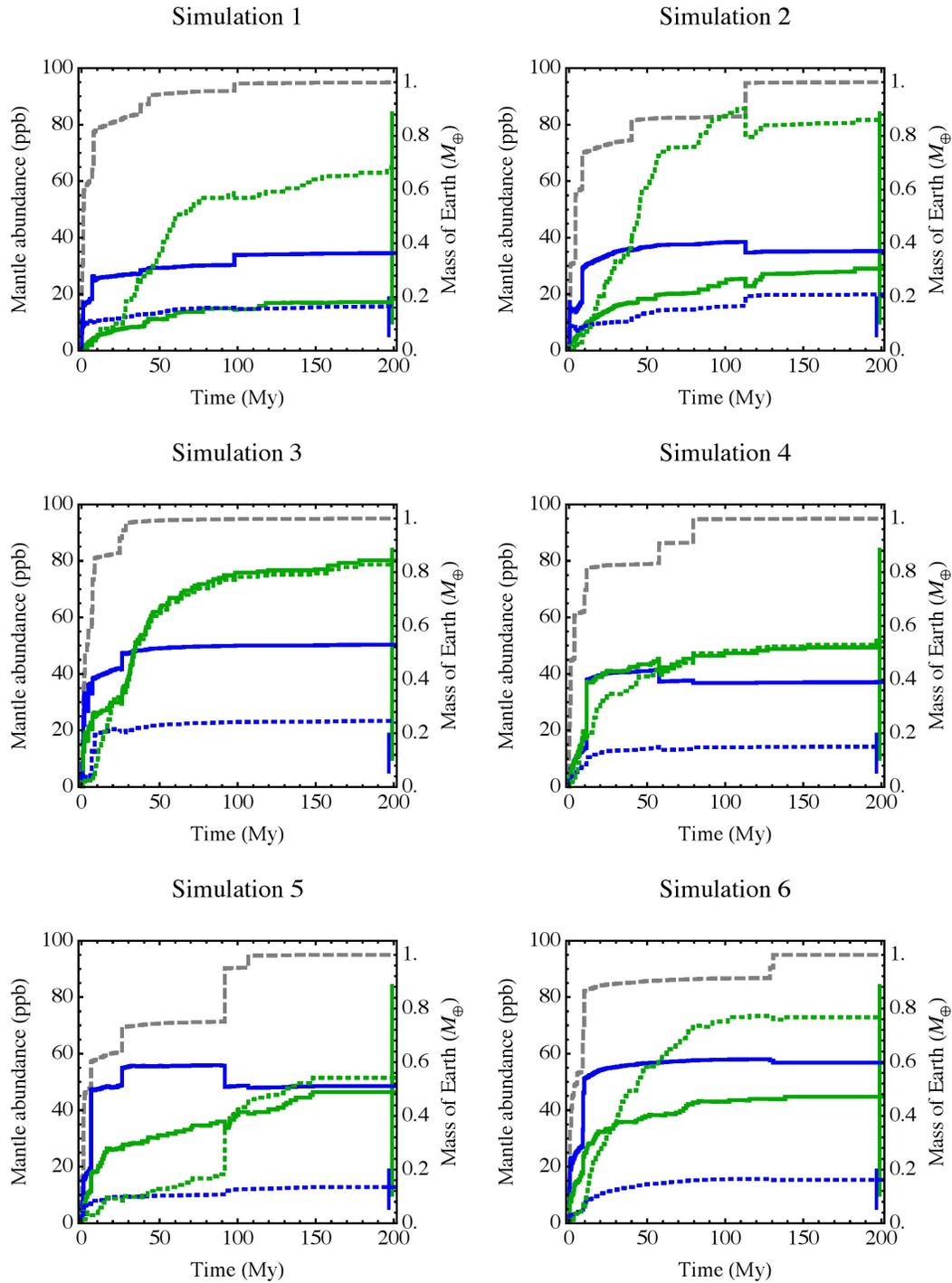

*Fig. 11: For each Earth-like planet in accretion simulations 1-6, the time evolution of the mantle abundances of W (blue) and Mo (green) is shown for fits to both the high (dashed) and low (solid) mantle C abundance estimates using the left-hand ordinate. The vertical bars on the right edge illustrate the two-sigma estimated W (blue) and Mo (green) abundance uncertainty for the BSE (Palme & O'Neill, 2014). For reference, the planetary growth curve (grey, dashed) for each Earth-*

*like planet is shown on the right-hand ordinate. Vertical jumps in Earth's growth curve indicate giant impacts whereas low slope sections show periods of slow growth from the accretion of planetesimals.*

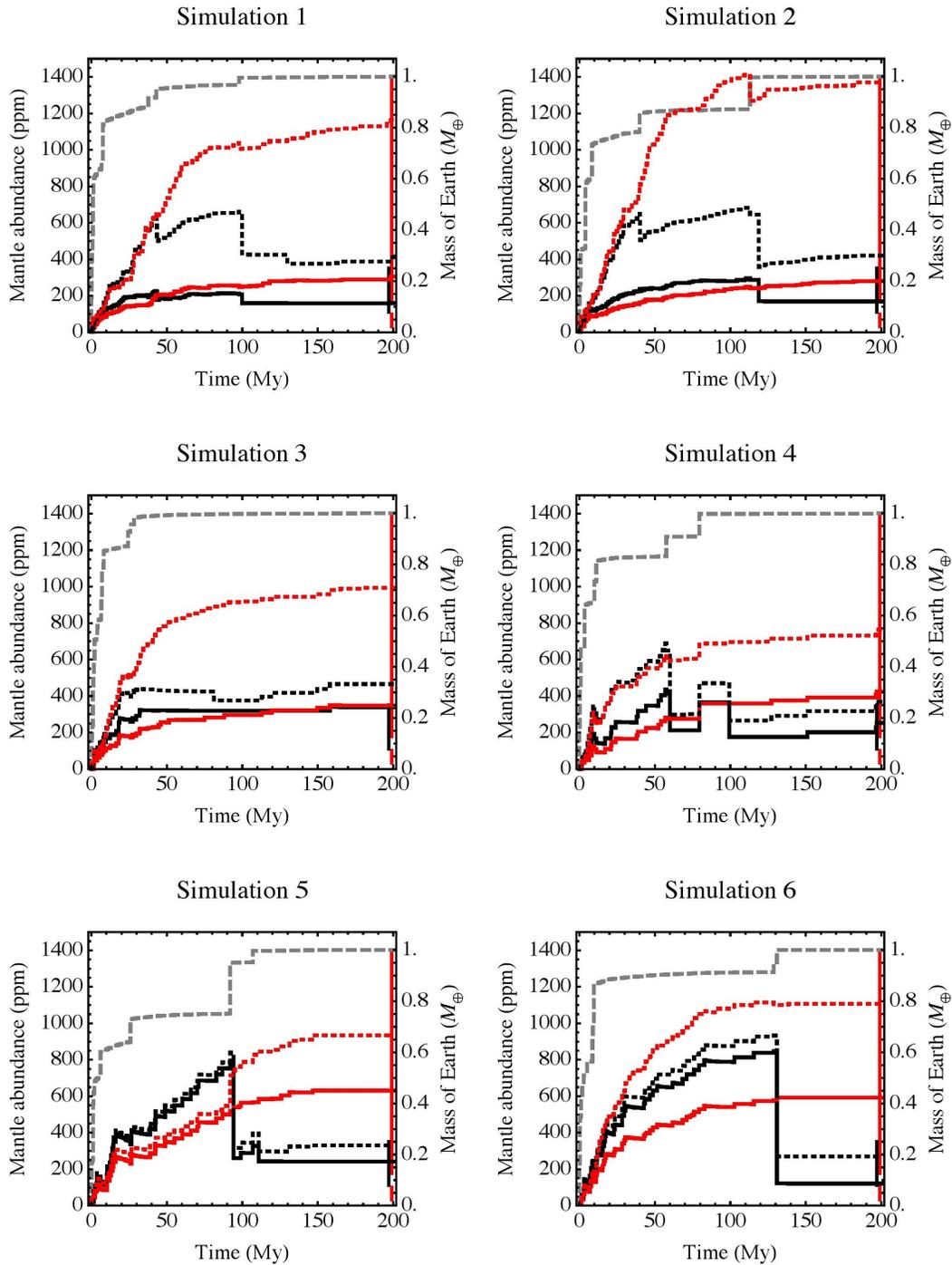

Fig. 12: For each Earth-like planet in accretion simulations 1-6, the time evolution of the mantle abundances of C (red) and S (black) is shown for fits to both the high (dashed) and low (solid) mantle C abundance estimates using the left-hand ordinate. The vertical bars on the right edge illustrate the estimated C (red) and S (black) abundance uncertainty for the BSE (Palme & O'Neill, 2014). For reference, the planetary growth curve (grey, dashed) for each Earth-like planet is shown on the right-hand ordinate.

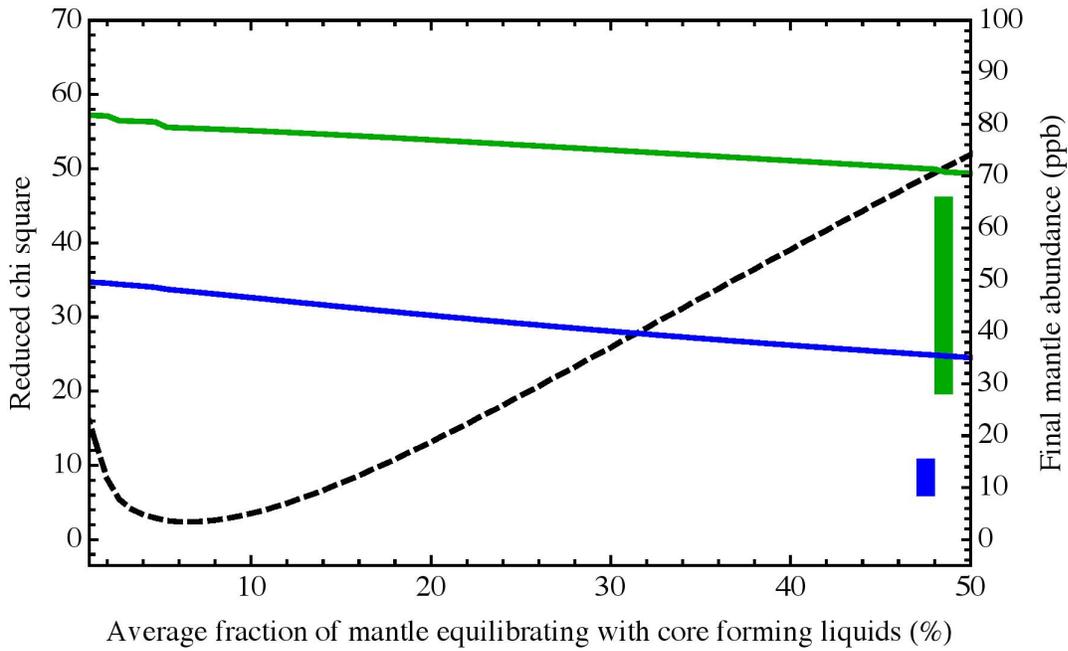

Fig. 13: The final mantle abundances of W (blue) and Mo (green) as a function of the average fraction of mantle equilibrating with core forming liquids after each giant (i.e. mantle magma ocean forming) and the overall reduced chi square metric (black, dashed) determined by fitting 17 measured element abundances in the BSE. The vertical bars near the right edge show the measured BSE abundances of W (blue) and Mo (green) with thickness given by 1σ uncertainties.

# Supplementary Information

## Partitioning of sulfur between metal and silicate

The partitioning of sulfur between liquid metal and silicate melt during each core formation event was calculated using the model of Boujibar et al. (2014, Eq. 11):

$$\log D_S^{met/sil} = \log X_{FeO}^{sil} - \log C_S + b/T + cP/T$$
$$+ d\log(1 - X_{Si}^{met}) + e(\log(1 - X_{Si}^{met}))^2$$
$$+ f(\log(1 - X_{Si}^{met}))^3 + g\log(1 - X_C^{met})$$
$$+ h\log(1 - X_{Fe}^{met}) + i\log(1 - X_{Ni}^{met})$$
$$+ j\log(1 - X_O^{met}) + k$$

In this equation, $D_S^{met/sil}$ is the metal-silicate partition coefficient of sulfur, $X_{FeO}^{sil}$ is the weight % concentration of FeO in the silicate melt, and the other terms are as defined in Boujibar et al. (2014). Rubie et al. (2016) used this equation but with the incorrect assumption that $X_{FeO}^{sil}$ is the mole fraction of FeO in the silicate melt. This mistake arose because (a) $X_{FeO}^{sil}$ in Eq. 11 of Boujibar et al. (2014) is simply defined as "concentration" and (b) in Eq. 6 of Boujibar et al. (2014) (on the same page as Eq. 11) $X_{FeO}^{sil}$ is defined as mole fraction.

The incorrect definition of $X_{FeO}^{sil}$ used by Rubie et al. (2016) resulted in the maximum concentrations of S in Earth's mantle during accretion being too high. Using the correct definition in the current paper causes the maximum concentrations to be lower by a factor of 3 to 4. However, the error in Rubie et al. (2016) has no effect on their results in terms of the final concentrations of S and the HSEs in Earth's mantle and core.

## Tungsten and molybdenum loss from the silicate mantle from pervasive metal saturation

We calculate the potential impact of a pervasive metal saturation event in a magma ocean using a simplified heterogeneous accretion model on mantle W and Mo concentrations.

High-pressure experiments have shown that $Fe^{3+}$ becomes increasingly stable relative to $Fe^{2+}$ at increasing pressures, which drives the disproportionation of FeO:

$3FeO = Fe_2O_3 + Fe$

This disproportionation may force the saturation of an immiscible Fe metal phase, and can be driven either by the crystallisation of $Fe^{3+}$-bearing aluminous bridgmanite in the lower mantle at pressures greater than around 26 GPa (Frost et al., 2003), or could occur in the liquid silicate at pressures of greater than around 23 GPa (Armstrong et al., 2019). Unlike metal added to Earth from impactor cores, which only equilibrate with a small fraction of Earth's silicate mantle, this pervasive precipitation event allows the complete equilibration of silicate with the fractionating metal throughout the mantle at the relevant high pressures.

For the sake of demonstrating the effect of such an event on W and Mo concentrations, we assume that this precipitation occurred throughout accretion from the point at which Earth was large enough for the deep mantle to be at pressures above 23 GPa threshold where this process may start to take place.

We have modelled the effect of iron precipitation on the tungsten concentration of the mantle using the same parametrisations for $K_D^{met-s}$ given in section 4.5 of the main text. For simplicity, we ignore activity effects and assume that W or Mo alloy with a pure Fe liquid. The model conceptually can be thought of as a growing planet, where a precipitation front at 23 GPa moves upwards with progressive accretion. We assume that a constant 0.5% metal precipitates from all parts of the silicate that are deeper than 23 GPa. log($fO_2$) increases linearly with fraction accreted to approximate the trend of impactors becoming more oxidised as accretion progresses, beginning at ΔIW-4 and ending at ΔIW-0.6, with an average ΔIW-2.3. We assume that precipitation occurs throughout the entire depth of the mantle to the CMB, rather than to the base of the magma ocean, because iron precipitation can occur within a crystal mush as well as a liquid, with a pyrolite liquidus pressure-temperature profile.

We find that 56% of mantle W is lost when Earth is fully accreted. This value is relatively insensitive to the choice of the pressure of the onset of precipitation (26 GPa gives 52% reduction), but decreases to 38% if the same overall amount of precipitation occurs at depths of 0.5*CMB to represent the magma ocean base. If 1% metal fractionates, mantle W loss increases to 62%, whereas if only 0.1% metal is lost, W loss decreases to 41%. The most sensitive parameter is $fO_2$, because of the high charge of the $W^{6+}$ cation. The amount of W lost increases to 74% if $fO_2$ evolution is not considered and a constant ΔIW-2.3 is assumed. Decreasing constant $fO_2$ by 1 log unit to ΔIW-3.3 increases this value to 92% and decreasing it by 1 log unit to ΔIW-1.3 dramatically decreases W loss to just 11%.

Mo follows a similar pattern of behaviour, but is less strongly sequestered by the precipitating metal (44% Mo is lost when Earth is fully accreted), and the $fO_2$-dependent response is dampened.

Whilst this model represents a significant simplification over reality because it does not couple iron precipitation with $fO_2$, it indicates that pervasive metal saturation is an efficient mechanism for the removal of W from the mantle (and, to a lesser extent, Mo). If pervasive metal saturation occurred, it provides a straightforward mechanism to reduce W and Mo concentrations.

**Supplementary Figures**

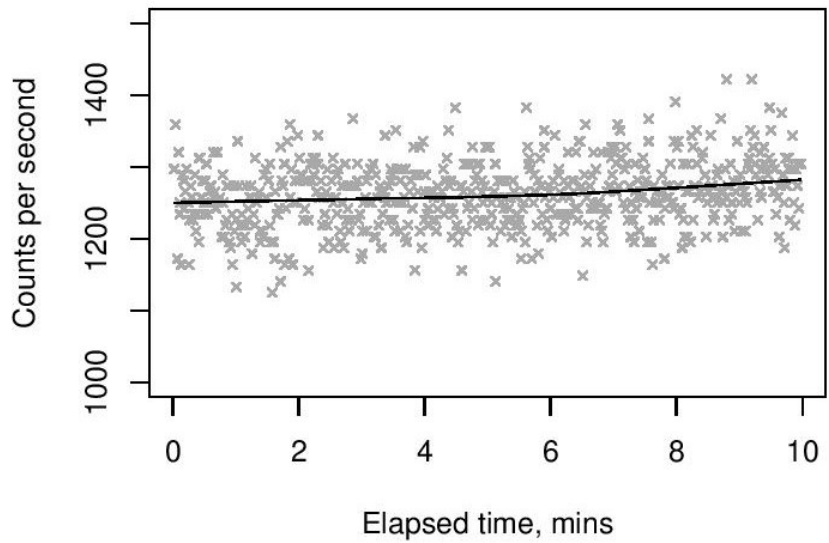

**Fig. S1** Time series showing apparent counts per second of Carbon measured on a pure Fe standard by EPMA on a single focussed spot. The observed drift, thought to result from C migrating towards the beam, is negligible over this time period.

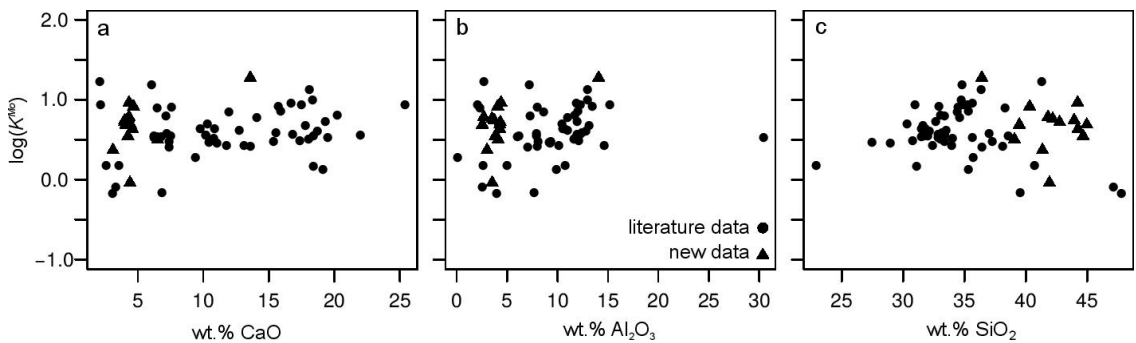

**Fig. S2** $\log(K^{Mo})$ as a function of a) CaO, with composition restricted in all components except CaO; b) $Al_2O_3$, with composition restricted in all components except $Al_2O_3$; c) $SiO_2$, where composition is restricted in all components except $SiO_2$ and MgO. Compositional restrictions are given in Table 4, where the limits are shown here by dotted lines. Circles are literature data and triangles are new data from this study.

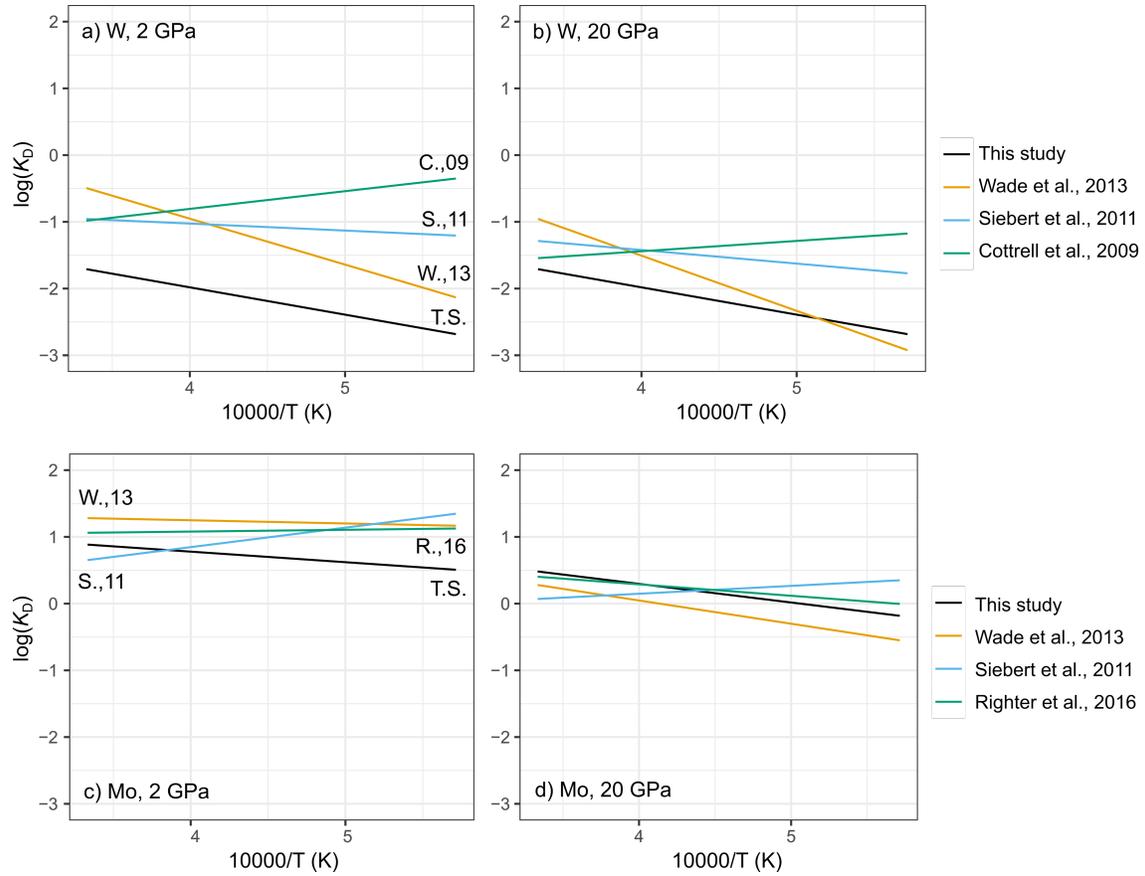

**Fig. S3** log($K_D^{Mo}$) and log($K_D^W$) calculated at 2GPa (a, c) and 20 GPa (b, d) from the predictive equations of our study and the previously published predictive expressions of Wade et al. (2013), Siebert et al. (2011), Cottrell et al. (2009) and Righter et al. (2016). Whilst all models give similar results, our model is most similar to that of Wade et al. (2013) in terms of temperature and pressure dependence. Notes regarding $K_D$ calculation: $\gamma_{Fe} = 1$ and infinite dilution of M in pure Fe assumed for published studies; median $\gamma_W$ (= 2.60) and $\gamma_{Mo}$ (=1.06) of experimental data used for this study; ideal mixing used for Siebert et al. (2011) following their method; NBO/T = 2.75 where required; molar chondritic mantle composition for Righter et al. (2016); $K_D^W$ recalculated to n = 6 ($WO_3$) for Cottrell et al. (2009) and Righter et al. (2011) for consistency, assuming ΔIW = -2.

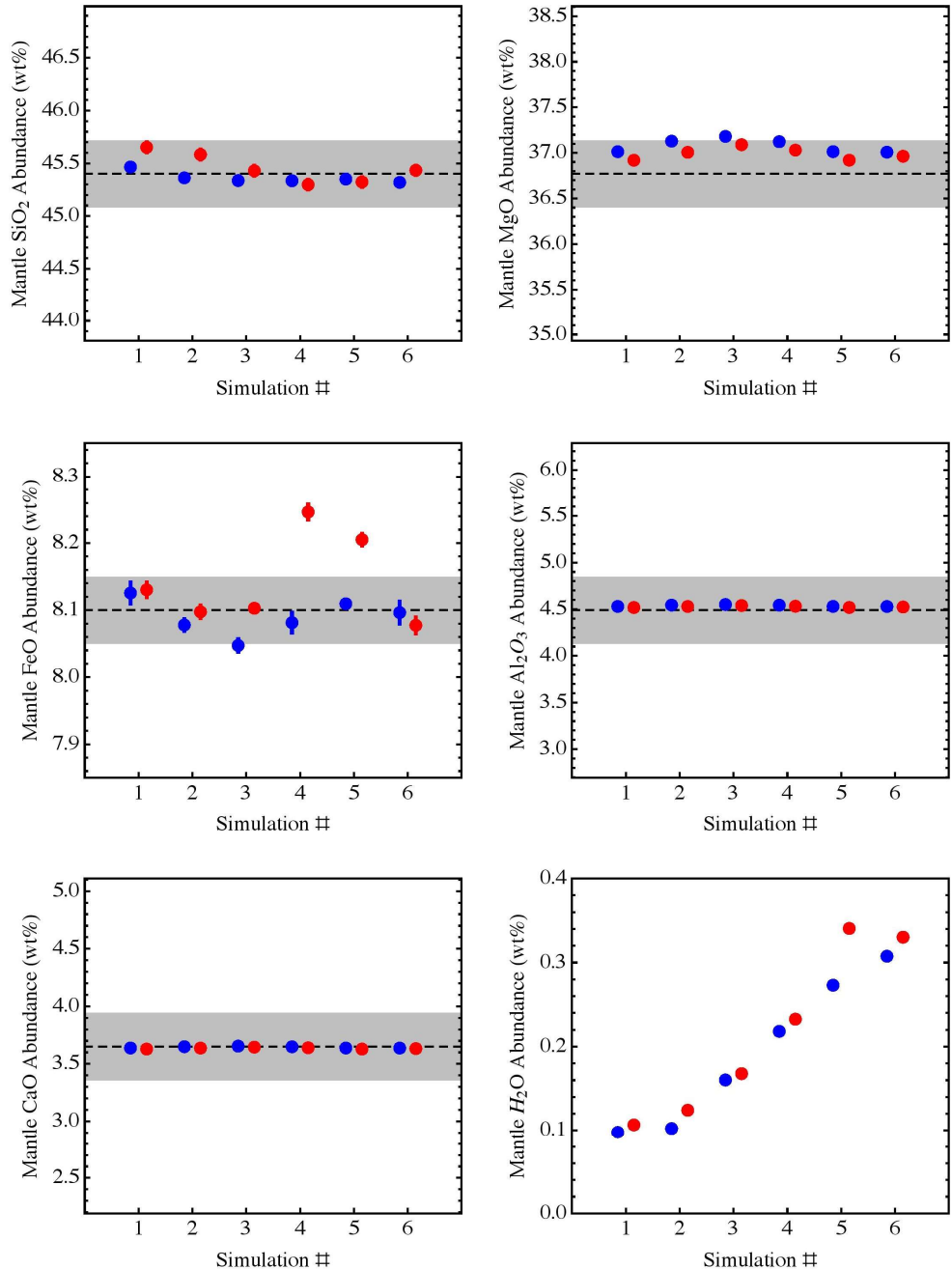

**Fig. S4** The best fit mantle abundances of major oxides ($SiO_2$, MgO, FeO, $Al_2O_3$, CaO, and $H_2O$) for the simulated Earth-like planets from simulations 1-6 when considering both a high (blue) and a low (red) C concentration estimate for Earth's mantle. Error bars of the best fit mantle uncertainties, when large enough to be visible, are the one-sigma laboratory measurement partition coefficient uncertainties propagated through the numerical model into absolute mantle abundances. The horizontal black dashed lined is the estimated mantle abundance for each major oxide with a corresponding shaded horizontal band indicating the one-sigma uncertainties of that estimate (Palme & O'Neill, 2014).

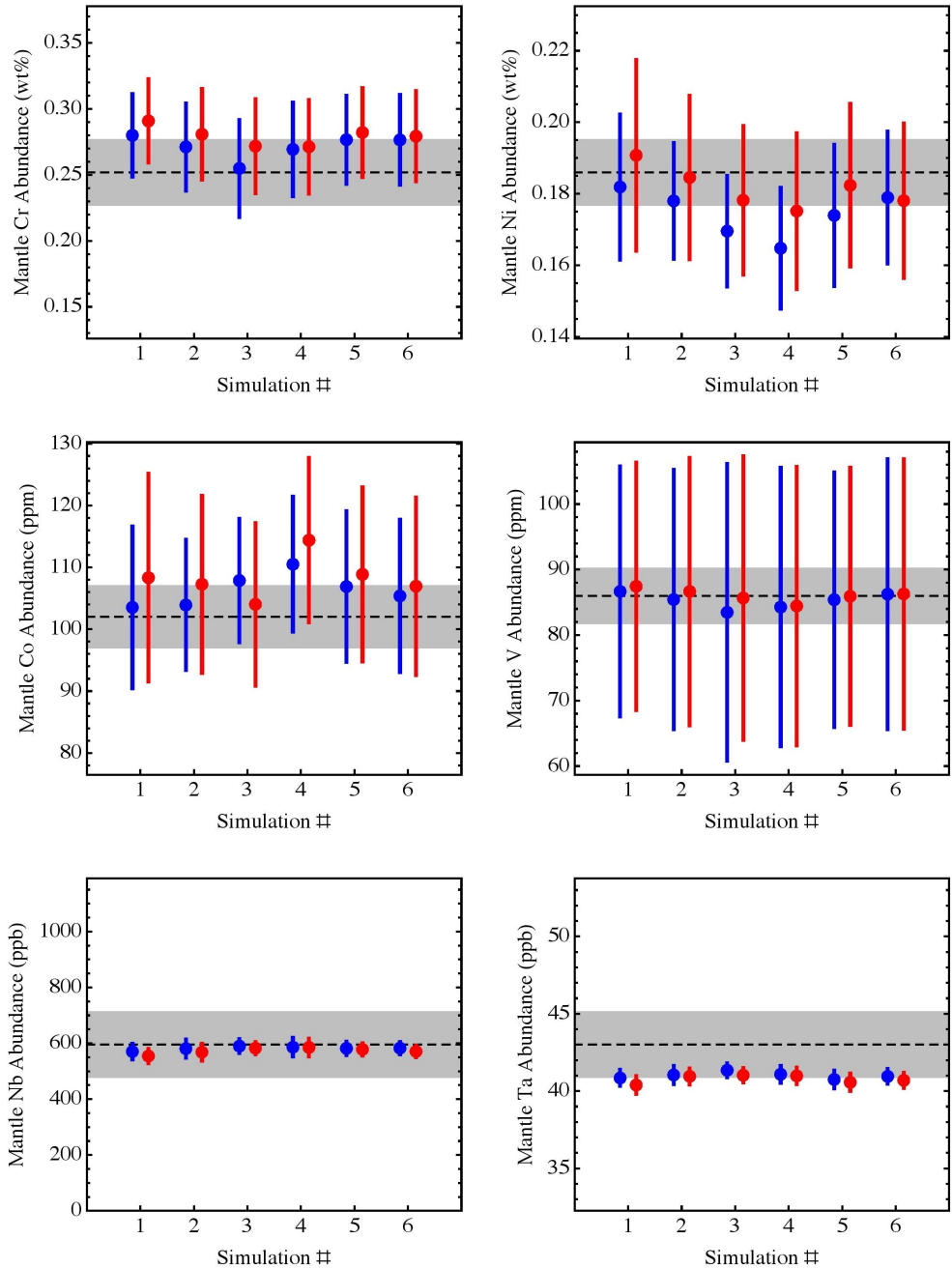

**Fig. S5** The best fit mantle abundances of refractory lithophile and moderately siderophile elements (Cr, Ni, Co, V, Nb, and Ta) for the simulated Earth-like planets from simulations 1-6 when considering both a high (blue) and a low (red) C concentration estimate for Earth's mantle. Error bars of the best fit mantle uncertainties, when large enough to be visible, are the one-sigma laboratory measurement partition coefficient uncertainties propagated through the numerical model into absolute mantle abundances. The horizontal black dashed lined is the estimated mantle abundance for each major oxide with a corresponding shaded horizontal band indicating the one-sigma uncertainties of that estimate (Palme & O'Neill, 2014).

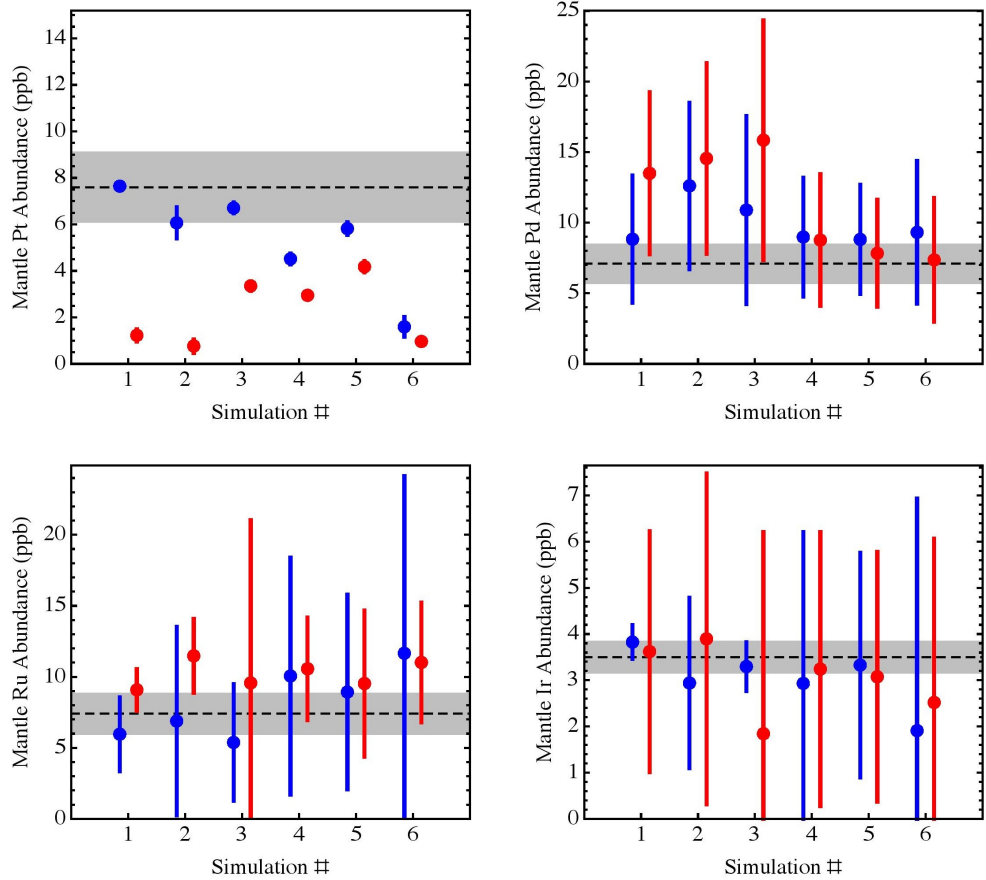

**Fig. S6** The best fit mantle abundances of highly siderophile elements (Pt, Pd, Ru, and Ir) for the simulated Earth-like planets from simulations 1-6 when considering both a high (blue) and a low (red) C concentration estimate for Earth's mantle. Error bars of the best fit mantle uncertainties, when large enough to be visible, are the one-sigma laboratory measurement partition coefficient uncertainties propagated through the numerical model into absolute mantle abundances. The horizontal black dashed lined is the estimated mantle abundance for each major oxide with a corresponding shaded horizontal band indicating the one-sigma uncertainties of that estimate (Palme & O'Neill, 2014).